\documentclass[12pt,epsfig]{article}
\usepackage{epsfig} 
\usepackage{latexsym,amssymb} 
\usepackage{amsmath} 

\DeclareMathAlphabet   {\mathsc}{OT1}{cmr}{m}{sc} 
 
\def\[{\left [} 
\def\]{\right ]} 
\def\({\left (} 
\def\){\right )}

\newcommand{\lbr}{\left\{} 
\newcommand{\rbr}{\right\}} 
\newcommand{\oline}[1]{\overline{#1}}

\newcommand{\wtd}[1]{\widetilde{#1}} 

\newcommand{\UV}       {\mathsc{uv}} 
\newcommand{\GS}       {\mathsc{gs}}

\newcommand{\GUT}      {\mathsc{gut}}

\newcommand{\hc}       {\mathrm{\; h.c. \;}}

\newcommand{\gappeq}{\mathrel{\rlap {\raise.5ex\hbox{$>$}} 
{\lower.5ex\hbox{$\sim$}}}} 
\newcommand{\lappeq}{\mathrel{\rlap{\raise.5ex\hbox{$<$}} 
{\lower.5ex\hbox{$\sim$}}}}

\begin{document}

\begin{flushright}
LPT Orsay--03/86 \\ FTUAM 03/25\\
IFT-UAM/CSIC--03--46 \\
hep-ph/0312155
\end{flushright}

\vspace{1.cm}

\begin{center}

{\Large {\bf Direct and Indirect Detection of Dark Matter in Heterotic Orbifold Models}}\\
\vspace{0.3cm}
{\large P. Bin\'etruy $^{1,3}$, Y. Mambrini $^2$, E. Nezri $^{1,3}$}\\
\vspace{0.3cm}
$^1$ Laboratoire de Physique Th\'eorique des Hautes Energies\\
Universit\'e Paris-Sud, F-91405 Orsay.\\
\vspace{0.2cm}
$^2$ Departamento de F\'isica Te\'orica C-XI and Instituto de F\'isica Te\'orica C-XVI,\\ 
Universidad Aut\'onoma de Madrid, Cantoblanco, 
28049 Madrid, Spain.\\
\vspace{0.2cm}
$^3$ APC , Universit\'e Paris 7,\\ Coll\`ege de France, F-75231 Paris 
Cedex 05.\\
\vspace{0.2cm} 
\end{center}

\abstract{We study the neutralino dark matter phenomenology in the context of 
effective field theories derived from the weakly--coupled heterotic string.
We consider in particular direct detection and indirect detection with 
neutrino telescopes rates. The two cases of moduli dominated and dilaton 
dominated SUSY breaking lead to completely different phenomenologies. 
Even if in both cases relic density constraints can be fulfilled, moduli
domination  generically leads to detection rates which are much below the
present and future experimental sensitivities,
whereas dilaton domination gives high detection rates accessible 
to the next generation of experiments. This could make dark matter searches an
alternative way to constrain high energy fundamental parameters. }

\newpage

\tableofcontents

\newpage

\section{Introduction}

There exists a large collection of measurements providing convincing evidence
in favor of the existence of cold dark matter in the universe \cite{Olive1}. 
But the exact nature of this dark matter  is still an open question.
One of the most promising and best motivated candidates is a Weakly 
Interacting Massive Particle (WIMP). 
Direct detection via its rare scattering with a nucleus in a detector,
 or indirect detection via its annihilation after 
gravitational storage in a massive body provide two possible experimental 
strategies.

  In the framework of supersymmetry (SUSY), most  extensions of the Standard 
Model (SM) predict a massive neutral weakly interacting particle in 
the form of a neutralino  ($ \chi^0_1  \equiv \chi$). Moreover, in the 
simplest versions of SUSY models such as the minimal supergravity model 
(mSUGRA),this particle is predicted to be stable. Recent works have 
constrained a large part of the parameter space available in mSUGRA  
\cite{mSUGRA} in constrained or unified versions of the model. 
For small values of $M_0$, 
we can even find in the literature some strong consequences on the limit of 
the neutralino mass : $< 500 $ GeV \cite{EllisWmap}  if we took into account 
all the recent accelerator analysis, but this can be strongly evaded when one
allows high values of $M_0$ and $m_{1/2}$ \cite{Nath1}. 
Up to now, several works have
generalized these simple models to a non--universal framework in the higgs 
sector \cite{EllisHiggs,Bottino1,Arnowitt1,Mynonuniv}, the gaugino sector 
\cite{Mynonuniv,BirkedalnonU,Nath2} or the sfermion sector \cite{Profumo:2003em}. Such analyses show that neutralino
dark matter searches are sensitive to the spectrum of supersymmetric 
particles, but the direct connection with the supersymmetry-breaking sector 
is not made because it appears only through the parameters of the  effective
theory, the so-called soft supersymmetry breaking parameters.

If one wants to make explicit the supersymmetry breaking mechanism, one
must identify the type of  breaking as well as the nature of the mediation
between the supersymmetry breaking sector and the observable sector. A standard
example is gaugino condensation in a hidden sector which interacts only 
gravitationally with  the sector of quarks and leptons. Since this involves 
gravitational interactions, it is natural to consider such a model in a string 
context and thus to include all fields associated with the gravitational 
sector, in particular the dilaton (whose vacuum expectation 
value determines the magnitude of the string coupling)
and moduli fields (whose expectation values determine the size of the 
compact manifold). Both types of fields play an important 
role in supersymmetry breaking.  The most elaborated class of models and 
probably the most realistic one from a phenomenological point of view 
is associated with the weakly coupled heterotic string  compactified on a
Calabi-Yau manifold or an orbifold.

Recently, the full one loop soft supersymmetry breaking terms in a large 
class of superstring effective theories have been calculated \cite{BiGaNe01} 
based on orbifold compactifications of the weakly--coupled heterotic string 
(including the so--called anomaly mediated contributions). Such models yield 
specific non-universalities which make their phenomenology significantly 
different from the minimal supergravity model. The parameter space 
in this class of models has already been severely constrained by taking into 
account accelerator and relic density constraints \cite{Bin1,Yanntalk} or benchmark 
models at the Tevatron \cite{NelsonTeva,Nelsontalk}.

In what follows, we take this specific class of models to discuss how direct 
and indirect neutralino detection depend on the properties of supersymmetry
breaking. Indeed, supersymmetry breaking induced by gaugino condensation 
may be transmitted through the auxiliary field vacuum expectation values of
the compactification moduli  or of the dilaton. We will see that the  two 
corresponding cases  have a very different behavior regarding dark matter
detection.

\section{Theoretical framework}

\subsection{Structure of heterotic orbifolds models at one loop}

The task of string phenomenology is
 to make contact between the high energy string theory, and the low energy 
 world. For this purpose, we need to build a superstring theory in four dimensions, 
 able to give us the Standard Model gauge group,  three generations of squarks,
 and a coherent mechanism of SUSY breaking. We will set here our analysis 
 in the framework of orbifold compactifications of the heterotic string within
 the context of the supergravity effective theory. We concentrate on those 
 models where the action is dominated by one loop order contributions to soft 
 breaking terms. The key property of such models is the non--universality of soft 
 terms, consequence of the beta--function appearing in the superconformal 
 anomalies. This  non--universality gives a particular phenomenology in 
 the gaugino and the scalar sector,
 modifying considerably the predictions coming from mSUGRA. In fact, these
 string--motivated models show new behavior that interpolates between the
 phenomenology of unified supergravity models (mSUGRA) and models dominated by
 the superconformal anomalies (AMSB). The constraints arising from accelerator
 searches and relic density have been already studied in \cite{Bin1}. It is thus interesting, to see to which extent direct and
 indirect detection of dark matter will be able to bring us extra information
 on the models.

We provide a phenomenological study within the context of orbifold compactifications of the weakly--coupled heterotic string, where we distinguish two regimes.
In the first one, the SUSY breaking is transmitted by the compactification
moduli $T^{\alpha}$, whose vacuum expectation values determine the size of the
compact manifold. Generic (0,2) orbifold models contain three $T_{\alpha}$ 
moduli fields. We considered a situation in which only an "overall modulus $T$"
field contributes to SUSY--breaking. The use of an overall modulus $T$ is equivalent
to the assumption that the three $T_{\alpha}$ fields of generic orbifold
models have similar contributions to SUSY--breaking. This is expected in the 
 absence of some dynamical effect that would strongly discriminate between the three moduli.
In our second example, it is the dilaton field $S$ present in any
four--dimensional string (whose vacuum
expectation value determine the magnitude of the unified coupling constant
$g_{\mathrm{STR}}$ at the string scale), that transmits, via its auxiliary
component, the SUSY breaking. We work in the context of models in
which string nonperturbative corrections to the K\"ahler potential act to
stabilize the dilaton in the presence of gaugino condensation 
\cite{BiGaWu96,BiGaWu97a}. The origin of the soft breaking terms are
 completely different in the two scenarii. Some are coming from the
 superconformal anomalies and are non--universal
 (proportional to the beta--function of the Standard Model gauge groups), 
others are generated in the hidden sector (from Green--Schwarz mechanism or
 gaugino condensation) and are thus universal. This mixture between
 universality and non--universality gives the richness of the phenomenology
 in this type of effective string models
 and confirms the interest of non--universal studies in the prospect of
 supersymmetric dark matter detection, the non--universality being 
in this case connected with the basic properties of the model.

\subsection{The moduli dominated scenario}

In the moduli dominated scenario, the one loop order supersymmetric SUSY
 breaking terms at GUT scale can be written \cite{BiGaNe01,GaNeWu99,GaNe00b}:

\begin{eqnarray}
M_a &=& \frac{g_{a}^{2}\(\mu\)}{2} \lbr 2
 \[ \frac{\delta_{\GS}}{16\pi^{2}} + b_{a}
\]G_2(T,\oline{T}) F^{T} + \frac{2}{3}b_{a}\oline{M} \rbr, \label{modsoftgaugi}\\
A_{ijk}&=& - \frac{1}{3} \gamma_{i}\oline{M} - p \gamma_{i} G_2(T,\oline{T})
F^{T} + {\rm cyclic}(ijk), \\ M_{i}^{2} &=& (1-p)\gamma_i
\frac{|M|^2}{9}. \label{modsoftscal}
\end{eqnarray}

\noindent
where $M_a$ and $M_i$ are the soft masses for
 the gauginos and scalars and $A_i$, the trilinear coupling.
$b_a$ is the beta--function coefficient for the gauge group $G_a$:

\begin{equation}
b_a=\frac{1}{16 \pi^2}
\left(
3 C_a - \sum_i C_a^i
\right).
\end{equation}

\noindent
where $C_a$, $C_a^i$ are the quadratic Casimir operators for the group
$G_a$ in the adjoint representation and in the representation of the
field $i$ respectively.
$F^S$ and $F^T$ are the auxiliary fields for the dilaton and the K\"ahler
modulus, respectively, $\overline{M}$ is the supergravity auxiliary fields
whose vacuum expectation value ($vev$) determines the gravitino mass
$m_{3/2}=-\frac{1}{3}\overline{M}$, and $\delta_{\mathrm GS}$ is
the Green--Schwarz coefficient which is a (negative) integer
between $0$ and $-90$. The function  
 $G_2(T, \overline{T})$ is proportional to the Eisenstein function and
vanishes when $T$ is stabilized at one of its two self--dual points.
From Eq.(\ref{modsoftgaugi}), it follows that when the moduli are stabilized at a self dual point, only
the second term contributes to gaugino masses. This is precisely the
"anomaly mediated" contribution.
The loop contributions have been computed using the Pauli--Villars (PV)
regularization procedure. The PV regular fields mimic the heavy string modes
that regulate the full string amplitude. The phenomenological parameter $p$
which represents the effective modular weight of the PV fields is constrained to be no larger than $1$, though it can be negative in
value. Thus the scalar squared mass for all matter fields is in
general non--zero and positive at one loop (only the Higgs can have a
negative running squared mass). The limiting case of $p=1$, where the
scalar masses are zero at one loop level and for which we recover a sequestered
sector limit, occurs when the regulating PV fields and the mass--generating
PV fields have the same dependence on the K\"ahler moduli. Another reasonable
possibility is that the PV masses are independent of the moduli, in which
case we would have $p=0$.
and $\gamma_i$ is related to the anomalous dimension through 
$\gamma_i^j=\gamma_i \delta_i^j$.
(see \cite{BiGaNe01,Bin1,NelsonTeva} for notations and conventions)

We clearly see in these formulae the competition between universal terms and
non--universal ones. The scalar mass terms are all non--universal 
and proportional to their anomalous dimension $\gamma_i$  and thus loop suppressed.
 The Green-Schwarz mechanism generates universal breaking terms for
the gauginos (proportional to $\delta_{\mathrm GS}$) whereas superconformal
 anomalies introduce non--universal contributions (proportional to $b_a$).
 The nature of the neutralino thus depends mainly on the value 
of the Green--Schwarz counterterm $\delta_{\mathrm GS}$, whereas the mass
 scale is the gravitino mass $m_{3/2}$.  

\subsection{The dilaton dominated scenario}

We turn now to a scenario where the dilaton is the primary source
of supersymmetry breaking in the observable sector. It is well known
that if we use the standard K\"ahler potential derived from the tree
level string theory, it is very difficult to stabilize the dilaton at
 acceptable weak--coupling values. We postulate in our study nonperturbative
correction of stringy origin to the dilaton K\"ahler potential. In that case, 
one condensate can stabilize the dilaton at weak coupling while simultaneously
ensuring vanishing expectation values at the minimum of the potential.
The key feature of such models is the deviation of the dilaton K\"ahler
metric from its tree level value. If we imagine the superpotential for the
 dilaton having the form $W(S) \propto e^{-3S/b_+}$, with
$b_+$ being the largest beta--function coefficient among the condensing
gauge groups of the hidden sector, then we are led to consider the
 phenomenology of models given by the following pattern of soft supersymmetry
 breaking terms \cite{BiGaNe01,GaNeWu99,GaNe00b}:

\begin{eqnarray}
M_{a}&=&\frac{g_{a}^{2}\(\mu\)}{2} \lbr  \frac{2}{3}b_{a}\oline{M}
+\[ 1 - 2 b_{a}' K_s \] F^{S} \rbr \label{dilatsoftgaugi}\\ A_{ijk} &=&
-\frac{K_s}{3}F^S - \frac{1}{3} \gamma_{i}\oline{M} +
\tilde{\gamma}_{i} F^{S} \lbr \ln(\mu_{\UV}^{2}/\mu_R^2)
-p\ln\[(t+\bar{t}) |\eta(t)|^4\] \rbr + (ijk) 
 \\ M_{i}^{2} &=& \frac{|M|^2}{9}
 \[ 1 + \gamma_i
-\(\sum_{a}\gamma_{i}^{a} -2\sum_{jk}\gamma_{i}^{jk}\) \(
\ln(\mu_{\UV}^{2}/\mu_R^2) -p\ln\[(t+\bar{t}) |\eta(t)|^4\] \) \]
\nonumber
 \\
 & &+ \lbr
 \wtd{\gamma}_{i}\frac{MF^S}{6}+\hc \rbr , \label{dilatsoftscal}
\end{eqnarray}

\noindent
where $\mu_{\UV}$ is an ultraviolet regularization scale (of the order of 
the string scale $M_{\mathrm{STR}}$) and $\mu_R$ the renormalization scale (taken at the
boundary value of $M_{\mathrm{GUT}}$)\footnote{For simplicity, we assume here
that $M_{\mathrm{STR}} \sim \mu_{\UV} \sim \mu_R \sim M_{\mathrm{GUT}}$
(the corresponding error is logarithmic and appears in a loop factor).}. Moreover

\begin{equation}
F^{S} = \sqrt{3} m_{3/2} (K_{s\bar{s}})^{-1/2}, ~~~~ 
K_{s\bar{s}} = \partial_s \partial_{\overline{s}} K .
\label{FS}
\end{equation}

\noindent
and

\begin{equation}
(K^{s\bar{s}})^{-1/2} = \sqrt{3}
\frac{\frac{2}{3}b_{+}}{1-\frac{2}{3}b_{+}K_{s}}, ~~~~
K_{s}=-g_{\mathrm{STR}}^2/2 . 
\label{Ktrue}
\end{equation}

\noindent
with $g_{\mathrm{STR}}$ the unified constant at $M_{\mathrm{STR}}$ 
(see \cite{BiGaNe01,Bin1,NelsonTeva} for the notations and conventions)
to ensure a vanishing vacuum energy in the dilaton--dominated limit.

The phenomenology of a dilaton--dominated scenario is completely different
from a moduli--dominated one. 
If we look at formulae (\ref{dilatsoftgaugi}) and (\ref{dilatsoftscal}),
 it is clear that we are in a domain of heavy squarks and sleptons 
(of the order of magnitude of the gravitino mass) 
and relatively light gauginos. 
Indeed, the factor $b_+$, as it contains a loop factor, can suppress
the magnitude of the auxiliary field $F^S$ relative to that of the 
supergravity auxiliary field $M$ through the relation (\ref{FS}). The
resulting gaugino soft breaking terms are less universal for low values 
of $b_+$.  

\section{Supersymmetric dark matter phenomenology}

\subsection{Relic density}

We recall that, in a general supersymmetric model, the neutralino mass matrix 
reads in the bino, wino, higgsino basis
($\tilde{B},\tilde{W}^3,\tilde{H}^0_1,\tilde{H}^0_2$) :
\begin{equation}
\arraycolsep=0.01in
{\cal M}_N=\left( \begin{array}{cccc}
M_1 & 0 & -m_Z\cos \beta \sin \theta_W^{} & m_Z\sin \beta \sin \theta_W^{}
\\
0 & M_2 & m_Z\cos \beta \cos \theta_W^{} & -m_Z\sin \beta \cos \theta_W^{}
\\
-m_Z\cos \beta \sin \theta_W^{} & m_Z\cos \beta \cos \theta_W^{} & 0 & -\mu
\\
m_Z\sin \beta \sin \theta_W^{} & -m_Z\sin \beta \cos \theta_W^{} & -\mu & 0
\end{array} \right)\;.
\label{eq:matchi}
\end{equation}
Thus the lightest neutralino $\chi$ is generically a superposition of these 
states :
\begin{equation}
\chi = z_{\chi,1} \tilde B+  z_{\chi,2}\tilde W  
+ z_{\chi,3}\tilde H_1  +z_{\chi,4} \tilde H_2
\label{eq:chi}
\end{equation}

In a large parameter space of the mSUGRA model, the $\chi$
is mainly bino--like because of the renormalization group evolution of
$m_{H_u}^2$ down to the electroweak scale. Indeed, let us recall
the radiative electroweak symmetry breaking  relation  
\begin{equation}
{\mu}^{2}=\frac{\(m_{H_d}^{2}+\delta m_{H_d}^{2}\) - 
  \(m_{H_u}^{2}+\delta m_{H_u}^{2}\) \tan^{2}{\beta}}{\tan^{2}{\beta}-1} 
-\frac{1}{2} M_{Z}^{2} ,
\label{radmuterm} 
\end{equation}
where $\delta m_{H_u}^2$ and $\delta m_{H_d}^2$ represent the one 
loop tadpole corrections to the running Higgs masses $m_{H_u}^2$ 
and $m_{H_d}^2$~\cite{ArNa92,BaBeOh94,PiBaMaZh97}. 
In  (\ref{radmuterm}) all the parameters are running and set at the 
minimization scale.
We see that, as $m_{H_u}^2$ is becoming negative
and relatively large in absolute value at low energy, $|\mu|$ which sets the 
higgsino mass scale, becomes large, in fact larger than the gaugino mass 
terms $M_1$ and $M_2$. Since in mSUGRA models, $M_{i=1,2}$
are linked at the weak scale by the  well-- known relation 
$M_1=\frac{5}{3} \tan^2 \theta_W M_2 \sim 0.5 M_2$ (with some corrections at
 2--loop order running), the $\chi$ aligns along its bino component. 
When $m^2_{H_u}$ (negative) decreases in absolute value, $\mu$ is then smaller,
 increasing the higgsino content of the neutralino.  In the $|m^2_{H_u}|$ 
decreasing direction, this leads successively to a mixed bino-higgsino neutralino and
 then an higgsino LSP and finally the no EWSB boundary when $\mu=0$. This
 happens typically in the hyperbolic branch (focus point) mSugra region thanks 
to the heavy scalar scale.

Different processes lead to a cosmologically favoured neutralino relic
density. For a bino neutralino one needs
light sfermions, $\tilde{\tau}(\tilde{t})$ coannihilation or
annihilation into pseudo-scalar $A$. If the lightest neutralino has a 
dominant wino component, the relic density drops down because of
efficient annihilations into gauge bosons as well as strong 
$\chi  \chi^+_1$  
coannihilations. For a non negligible higgsino component 
 neutralino annihilates into gauge bosons or $t\bar{t}$
and relic density is also decreased by $\chi\chi^+_1$ and
$\chi\chi^0_2$ coannihilations.

\subsection {Direct detection}

Direct detection consists of measuring the energy deposited in a low 
background detector by the recoil 
of a nucleus from its elastic scattering with a Weakly Massive Interacting
Particle (WIMP) \cite{Goodman}. 
In our case, the best WIMP candidate is the lightest neutralino $\chi$. The rates follow the neutralino-proton
spin--independent elastic cross sections ($\sigma^{scal}_{\chi-p}$) or
spin--dependent one ($\sigma^{spin}_{\chi-p})$,
function on the target nucleus spin \cite{Jungman,Munoz:2003gx}.

$\sigma^{scal}_{\chi-p}$ is essentially driven by first generation scalar quark 
($\tilde u_i$, $\tilde d_i$) exchanges 
or neutral Higgs ($h$, $H$) 
($\chi q \xrightarrow{H,\tilde{q}} \chi q$) and the spin--dependent 
one $\sigma^{spin}_{\chi-p}$ by first generation squark and $Z$
exchanges ($\chi q \xrightarrow{Z,\tilde{q}} \chi q$). The processes involving
$Z$-boson exchange being 
completely dependent on the neutralino higgsino fraction.

The main impact of the null searches, particularly at LEP, is in the increase
in the lower limit to the LSP mass, as well as the rest of the sparticle 
spectrum.

 In the mSUGRA case, the bino-like nature of the lightest neutralino $\chi$
implies a highly suppressed scalar
cross section via heavy neutral higgs exchange because of the low couplings
$\chi H \chi$ (proportional to the product of their higgsino
and gaugino component) at moderate value of $\tan \beta$. 
If we look at the spin--dependent cross section, dominated by $Z$
exchange, it becomes much larger when $\chi$ is mostly higgsino because 
of the enhancement due to the $\chi Z \chi$
coupling (proportional to the square of the higgsino components).

The real possibilities
of an enhancement of  $\sigma^{scal}_{\chi-p}$ are thus in the high 
$\tan \beta$ regime or in models that  predict low first generation squarks 
masses for moderate gaugino masses, like some of the string inspired models 
that we study in this paper.
Another possibility is to find regions of parameter space that increase the 
higgsino component of the lightest neutralino, enhancing in the same way    
its coupling.  This happens for large $M_0$ in mSUGRA along the ``no EWSB''
boundary. We will see that for a large class of heterotic
orbifold models, all these constraints can be achieved. At the opposite, 
some of them will exhibit a complete depletion of the scalar cross--section, 
far below the sensitivity of the next generation of detectors. In this sense,
direct detection of dark matter can become an important tool in the
effort of discriminating string inspired models and constraining  their 
parameter space.

The current experimental status may be briefly summarized as follows.
Although  one of the current experiments, the DAMA collaboration 
\cite{Dama} claimed an evidence and gave a determination of the allowed 
maximum--likehood region in the WIMP--mass and WIMP--nucleon cross
section of $10^{-6}-10^{-5}$pb for a WIMP's mass between 30 and 270 GeV,
other experiments exclude almost (CDMS \cite{Cdms}, EDELWEISS
\cite{Edelweiss}) or completely (ZEPLIN I \cite{Zeplin}) the DAMA region. 
But many new or upgraded versions of direct
detection experiments will soon reach a significantly improved sensitivity
for WIMP detection (EDELWEISS II \cite{EdelweissII}, ZEPLIN(s) 
\cite{Zeplin}). Our study will be
placed in the light of this next generation of detectors to see whether 
a SUSY dark matter candidate could be directly detected in the next years,
and how this would constrain some of the fundamental SUSY breaking terms, 
and, as a consequence, the parameters of the more fundamental string theory.

\subsection {Neutrino indirect detection}

Neutralinos can also be gravitationally captured in massive
astrophysical bodies like the Sun or the Earth by successive diffusions on 
their nuclei leading to a trapped neutralino population at their center. Then
neutralinos can annihilate and the annihilation products,
essentially gauge bosons and heavy quarks, decay emitting neutrinos. 
After conversion into muons through the Earth, these neutrinos can be 
observed by neutrino telescopes collecting \v{C}erenkov light
of induced muons traveling in water or in ice. The  annihilation rate
depends both on the capture rate ($\sigma^{scal/spin}_{\chi-p}$ depending on
the target nucleus spin) and
on the neutralino annihilation cross section \cite{Jungman}. 
The capture rate in the Earth depends on $\sigma^{scal}_{\chi-p}$
because of the zero spin of the iron nucleus. This leads to muon fluxes
far beyond reach of detection \cite{Myuniv}. In the case of the Sun, capture 
rates are enhanced by $Z$ exchange in $\sigma^{spin}_{\chi-p}$ thanks to 
the non zero spin of the hydrogen. Fluxes are then maximized for a substantial 
higgsino fraction. Furthermore, mixed higgsino neutralino states
annihilate into $W^+W^-,\ ZZ$ or $t\bar{t}$ leading to more energetic 
neutrinos/muons than other annihilation channels. This happens along the 
``noEWSB'' boundary, where (thanks to $m^2_{H_u}$ running) the neutralino gets 
a dominant higgsino fraction. This is the case in mSUGRA for high values of 
$M_0$ \cite{Feng,Myuniv} and can be strongly
favored with non universal gaugino masses by decreasing $M_3|_{GUT}$ 
\cite{Mynonuniv}.

Some experiments have constrained these fluxes (Macro \cite{Macro}, Baksan 
\cite{Suvorova}, Super-Kamiokande \cite{SuperK}), but future neutrinos 
telescopes like Antares \cite{AntarLee} and Icecube \cite{Ice3Edsjo} will be
 much more efficient. They will improve current sensitivities of order 
${\rm 5.10^3 \mu\ km^{-2}\ yr^{-1}}$ to 
${\rm 10^3-10^2\ \mu\ km^{-2}\ yr^{-1}}$ on muon fluxes coming from 
neutralino annihilations in the centre of the Sun. We will compare our 
predictions for fluxes coming from the Sun with both current and future 
sensitivities.

As we will see below, in the models that we are considering in this work, the
muon fluxes coming from the Sun are very  dependent on the nature of 
SUSY breaking, i.e. whether it is dominated by moduli or dilaton $F$ terms. 
In this way, a neutralino dark matter signal in a neutrino telescope would 
provide key information on the nature of SUSY breaking and on the fundamental 
underlying theory.

\vskip 0.2cm

Works on prospects for direct and/or neutrino indirect
detection of neutralino dark matter in mSUGRA/CMSSM or non universal 
frameworks can be found in \cite{Nath1,EllisHiggs,Ellis:2003ry,Bottino1,Mynonuniv,BirkedalnonU,Chattopadhyay:2003yk,Cerdeno:2003yt,Baer:2003jb,Hooper:2003ka,Feng,Myuniv,Profumo:2003em}.

\subsection{Tools and experimental constraints}

\label{sec:constraints}

In this section we describe the tools that we have used and the various 
constraints that we have imposed to 
obtain a correct phenomenology at the electroweak scale for our analysis.

We have used the Fortran code {\tt SuSpect2}~\cite{Suspect} to solve the  
renormalization group equations (RGEs) for the soft supersymmetry breaking 
parameters between the high energy boundary scale $\mu_{\UV}$ and the 
scale given by the Z-boson mass (electroweak scale). While the
initial scale $\mu_{\UV}$ should itself be treated as a
model-dependent parameter, for our purposes we have chosen for
$\mu_{\UV}$ the scale of grand unification $M_{\GUT}$. We use $\tan\beta$ 
and the sign of the supersymmetric $\mu$ parameter in the superpotential
as free parameters, defined at the low-energy (electroweak) scale.

The magnitude of the $\mu$ parameter is determined by imposing
electroweak symmetry breaking (EWSB) at the usual scale 
$(m_{\tilde t_1} m_{\tilde t_2})^{1/2}$ ~\cite{GaRiZw90,deCa93}. 
The one-loop corrected $\mu$ is obtained from the condition (\ref{radmuterm}).

The soft supersymmetry breaking parameters at the weak 
scale are then passed on to the C code {\tt micrOMEGAs}~\cite{Micromegas} 
to perform the calculation of physical masses for the superpartners and 
various indirect constraints, to be described below. 

We note that
the value of the $\mu$ parameter is fundamental in the analysis of 
astroparticle processes, because it determines the nature, mass $and$ couplings
of the lightest neutralino $\chi$ which we require to be the LSP. We then 
estimated the detection rates using the {\tt DarkSusy} package \cite{Darksusy}.
 
The remaining parameter space is further reduced by limits on 
superpartner and Higgs masses from various collider experiments. 
We take the most recent bounds given by the different experiments of 
the LEP Working Group~ \cite{LHWG01,LHWG02}. Concerning 
the light CP-even neutral Higgs mass ($m_h$), we assume that a 95\% 
confidence level (CL) lower limit on $m_h$ is set at 111.5 GeV. 
The search for an invisibly decaying 
Higgs boson in $hZ$ production has allowed a 95\% CL lower limit 
on $m_h$ to be set at 114.5 GeV, assuming a production 
cross section equal to that in the Standard Model and a 100\% 
branching fraction to invisible decays~\cite{Higgslimit}. 
We believe the value of 113.5 GeV will serve as a good 
mean. Concerning the chargino limit, we take 103.5 GeV, 
bearing in mind that in some degenerate cases and for 
light sleptons the limit can go down to 88 
GeV~\cite{charginolimit}. For the squark sector the limit of 97 
GeV~\cite{Stoplimit} is used. For all mass 
bounds we should keep in mind that experimental limits 
are always given in the context of a particular SUSY 
model framework which is not generally a string motivated 
one. The bounds we use could possibly be weakened in some 
cases.

Various non-collider observations can be used to further reduce 
the allowed parameter space of the loop-dominated orbifold models that we 
consider. We will focus our attention on the three sets of data 
that are the most constraining for these models: the density of relic 
neutralino LSPs, the branching ratio for decays 
involving the process $b \to s \gamma$ and the measurement of the 
anomalous magnetic moment of the muon. We apply :

{\bf Relic density:}

Recent evidence suggests~\cite{Pr00} that $\Omega_{
\chi} \sim 0.3$ with ${\rm h}^2 \sim 0.5$. We will take as a 
conservative favored region

\begin{equation} 
0.03<\Omega_{\chi}{\rm h}^2<0.3. \label{limomega} 
\end{equation}

In addition to our 
conservative dark matter limit, we also take into account the 
recent results of WMAP~\cite{WMAP} 
that give a 2$\sigma$ 
range for the density of cold dark matter, $\Omega_{\rm CDM} {\rm 
h}^2 = 0.1126^{+0.0161}_{-0.0181}.$ 
Let us stress that the 
requirement of~(\ref{limomega}) should not be treated as an exact 
constraint, but rather as an indication of the region preferred by 
cosmological considerations because of the uncertainty in mass spectrum
 calculation.

{\bf $ b \rightarrow s \gamma$ Constraint:} 

Another observable where the SUSY particle contributions can be 
important and measurable is the flavor changing decay $ b 
\rightarrow s \gamma$~\cite{Bertolini}. In the Standard Model, 
this process is mediated by virtual isospin $+1/2$ quarks and 
$W$-bosons. In supersymmetric theories, the spectrum allows new 
contributions involving loops of charginos and squarks or top 
quarks and charged Higgs bosons. For our 
analysis, we use the results given by the CLEO and BELLE 
collaborations~\cite{Cleo}. 
We adopt the procedure taken in 
the recent benchmark study of Battaglia et al.~\cite{bench} and 
choose to impose the constraint 
\begin{equation} 
2.33 \times 10^{-4} < \mathrm{BR} (b \rightarrow s \gamma) < 4.15 
\times 10^{-4} . 
\end{equation}

{\bf The Muon Anomalous Magnetic Moment:}

Recently, the Brookhaven collaboration has given a new measurement 
of the anomalous magnetic moment of the muon~\cite{Brown} 

\begin{equation} 
\frac{(g^{\rm exp}_{\mu}-2)}{2} = a_{\mu}^{\mathrm{exp}} 
=11~659~202~(14)~(6)\times 10^{-10}, 
\label{muonexp} \end{equation} 

\noindent
Following~\cite{MaWe03} we introduce the parameter $\delta_{\mu}$ 
to quantify the difference between theoretical and experimental 
determinations of $a_{\mu}$: 

\begin{equation} 
\delta_{\mu} \equiv (a_{\mu} - 11~659~000 \times 10^{-10})\times 
10^{10} . 
\end{equation} 

\noindent
From this the current 
experimental determination of the parameter $\delta_{\mu}$ is 
$\delta_{\mu}^{\mathrm{exp}}= 203 \pm 8$. 
In our discussion, we are less conservative than 
the authors of~\cite{MaWe03} and consider a 2 standard deviation 
region about the anomalous moment of the muon based on the $\tau$ 
decay analysis \cite{Davier}:

\begin{equation}
 -11.6~<~\delta_{\mu}^{\mathrm{new \, physics}} = 
 \delta_{\mu}^{\mathrm{exp}}-\delta_{\mu}^{\mathrm{SM}}~<~30.4 ~~~~~
 [2~\sigma]. 
\end{equation}

\vskip .5cm

\section{The Models}

\subsection{The general moduli-dominated case}

\begin{figure}

    \begin{center}

\centerline{
       \epsfig{file=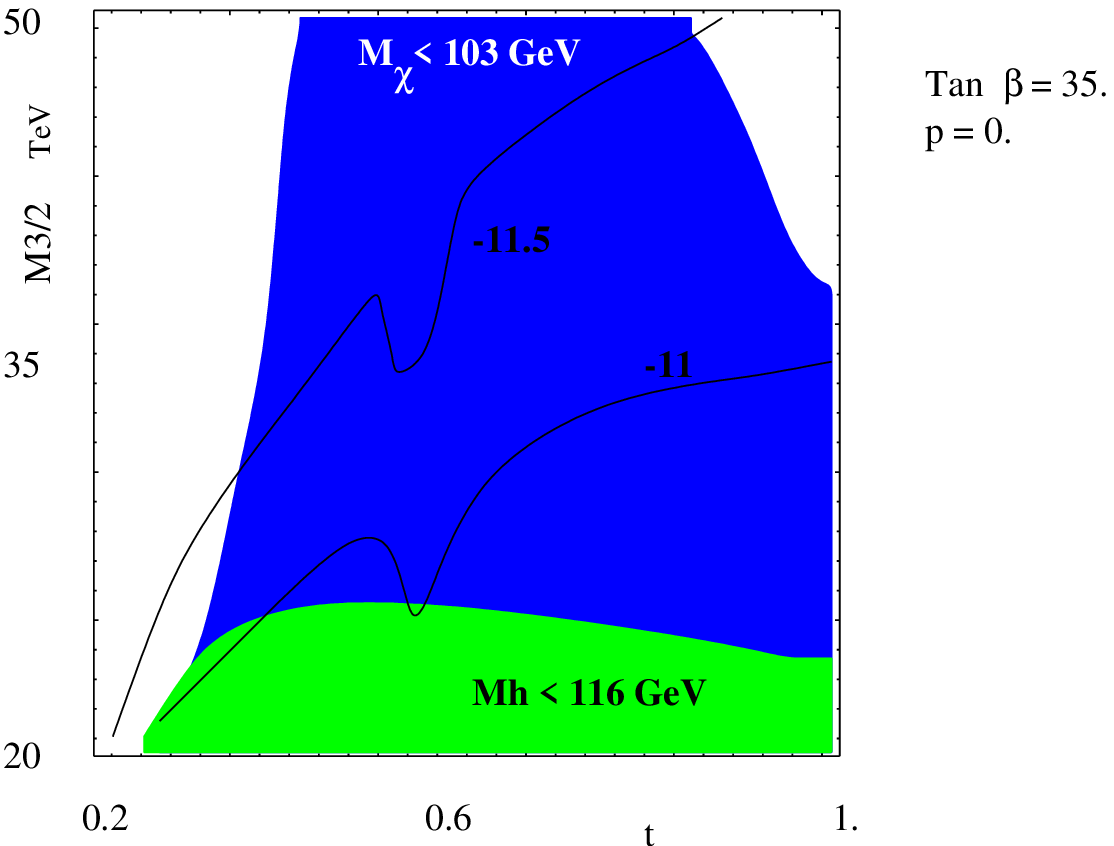,width=0.45\textwidth}
       \epsfig{file=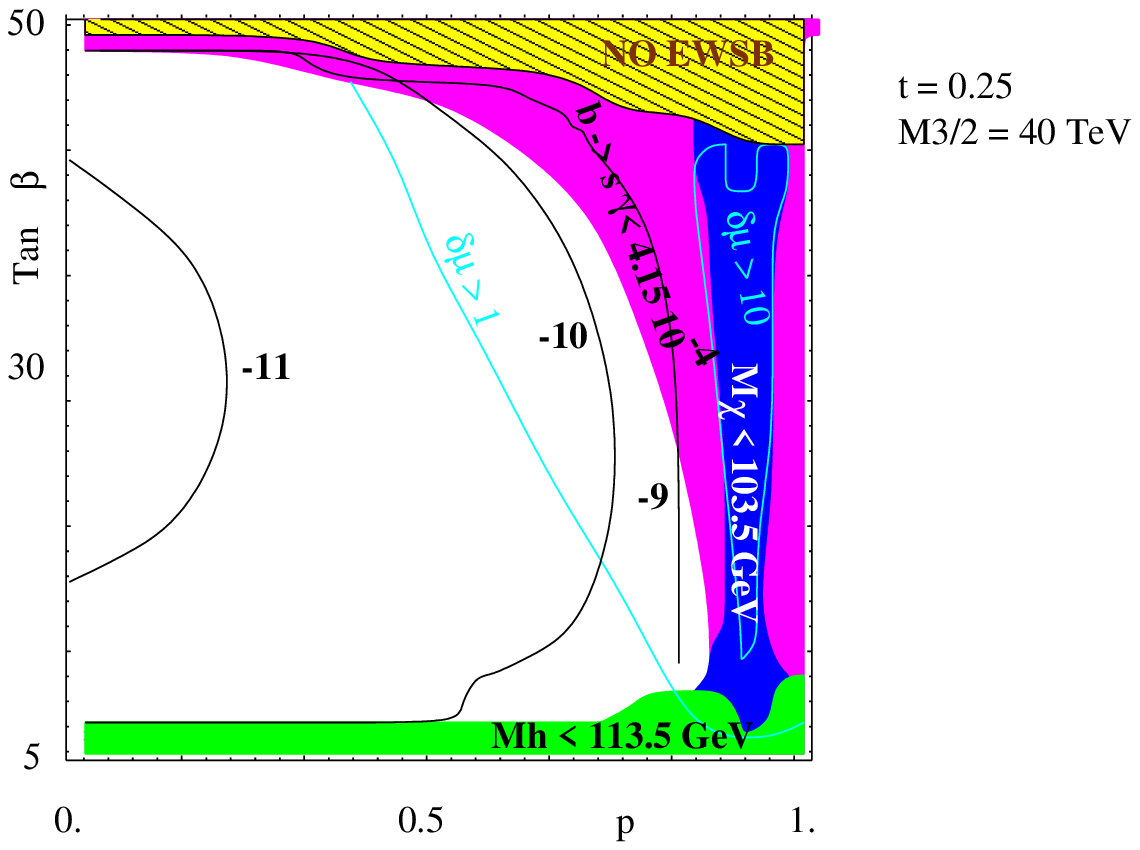,width=0.46\textwidth}}
          \caption{{\footnotesize {\bf The spin--independent scalar
          cross section in the moduli
          parameter space for $\delta_{GS}=p=0$}, in the ($t$, $M_{3/2}$)
          plane, for tan $\beta$ $=35$ (left) and ($p$, tan$\beta$) plane for
          $M_{3/2}=40$ TeV (right). Accelerators and
          cosmological constraints are given for $\mu > 0$. 
          The labels on the black lines correspond to the $Log_{10}$ value
          of $\sigma^{scal}_{\chi-p}$ (pb).
          For a description of
          the experimental constraints applied, see 
          Section~\ref{sec:constraints}.}}
        \label{fig:dirmoduli_p}
    \end{center}
\end{figure}

\begin{figure}
    \begin{center}
\centerline{
       \epsfig{file=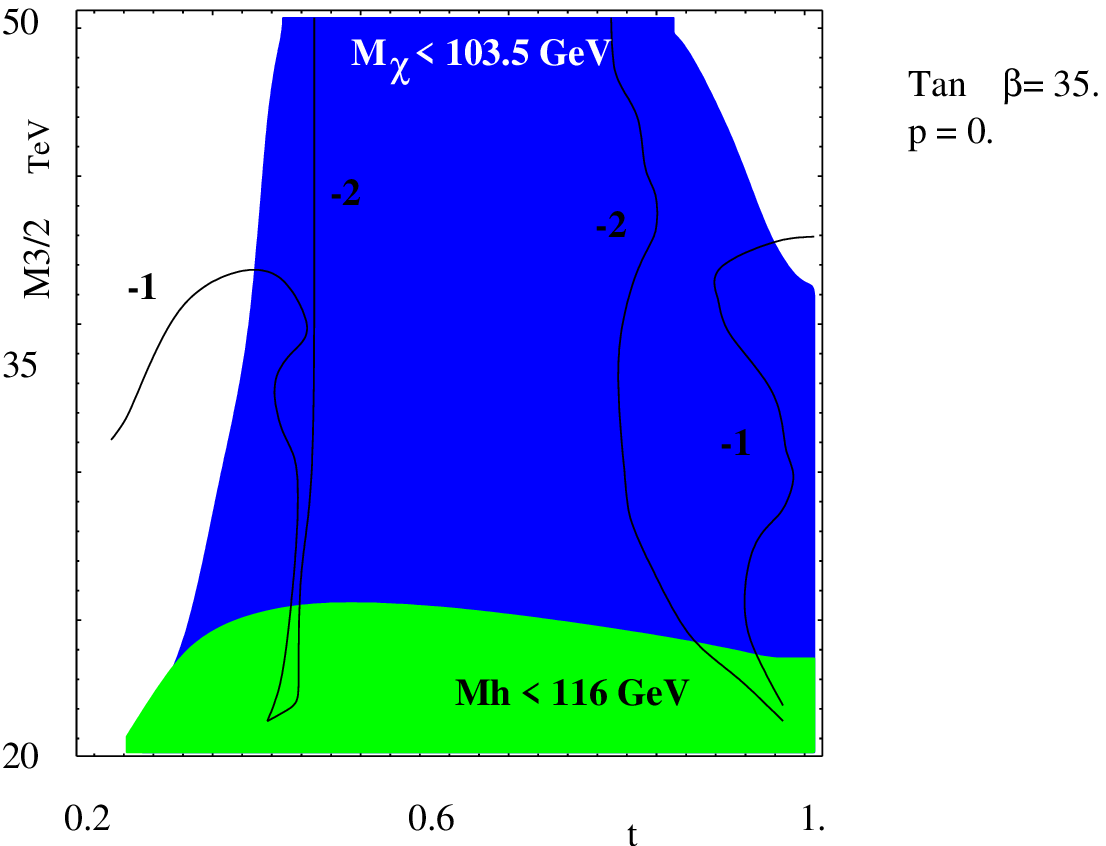,width=0.45\textwidth}
       \epsfig{file=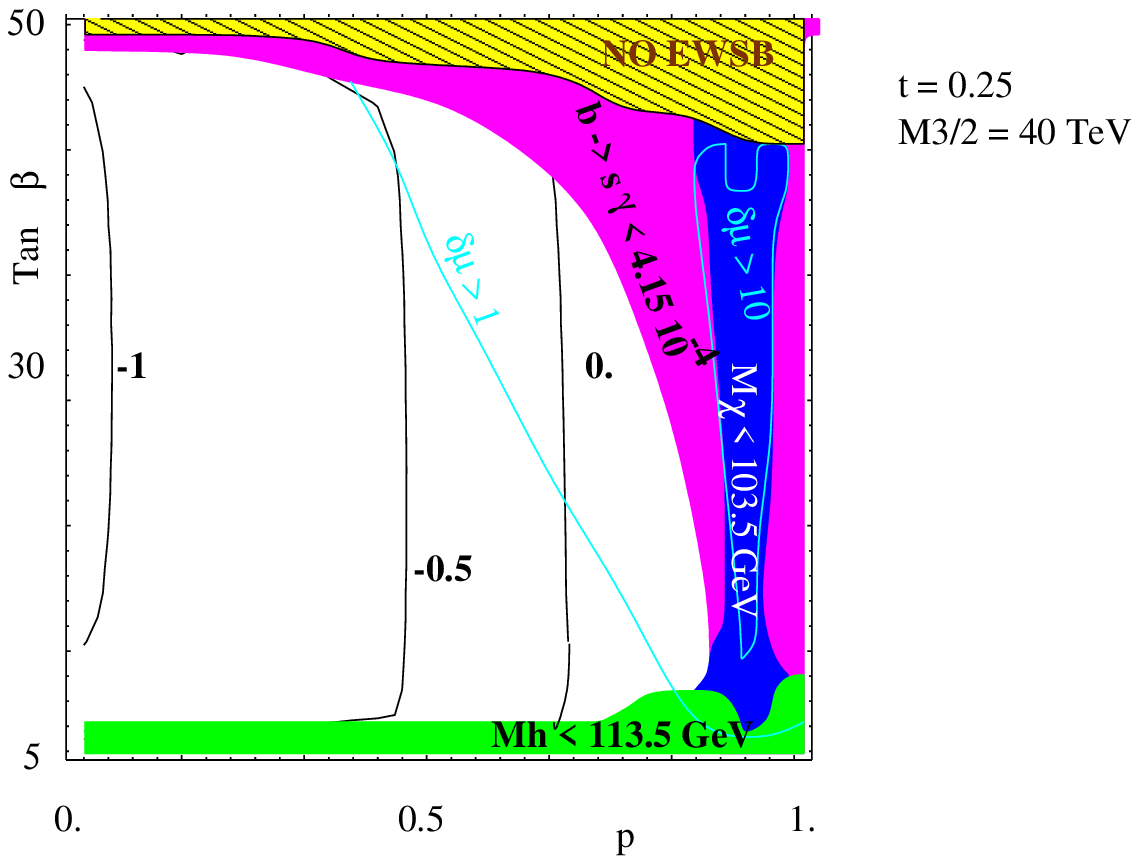,width=0.46\textwidth}}
          \caption{{\footnotesize {\bf Muon fluxes from the sun in the moduli
          parameter space for $\delta_{GS}=p=0$}, in the ($p$, $M_{3/2}$)
          plane, for tan $\beta$ $=35$ (left) and ($t$, tan$\beta$) plane for
          $M_{3/2}=40$ TeV (right). Accelerators and
          cosmological constraints are given for $\mu > 0$. 
          The labels on the black lines correspond to the $Log_{10}$ value
          of the flux ($ \mu\ {\rm km^{-2}\ yr^{-1}}$).
          For a description of
          the experimental constraints applied, see 
          Section~\ref{sec:constraints}.}}
        \label{fig:indmoduli_p}
    \end{center}
\end{figure}

\begin{figure}
    \begin{center}
\centerline{
       \epsfig{file=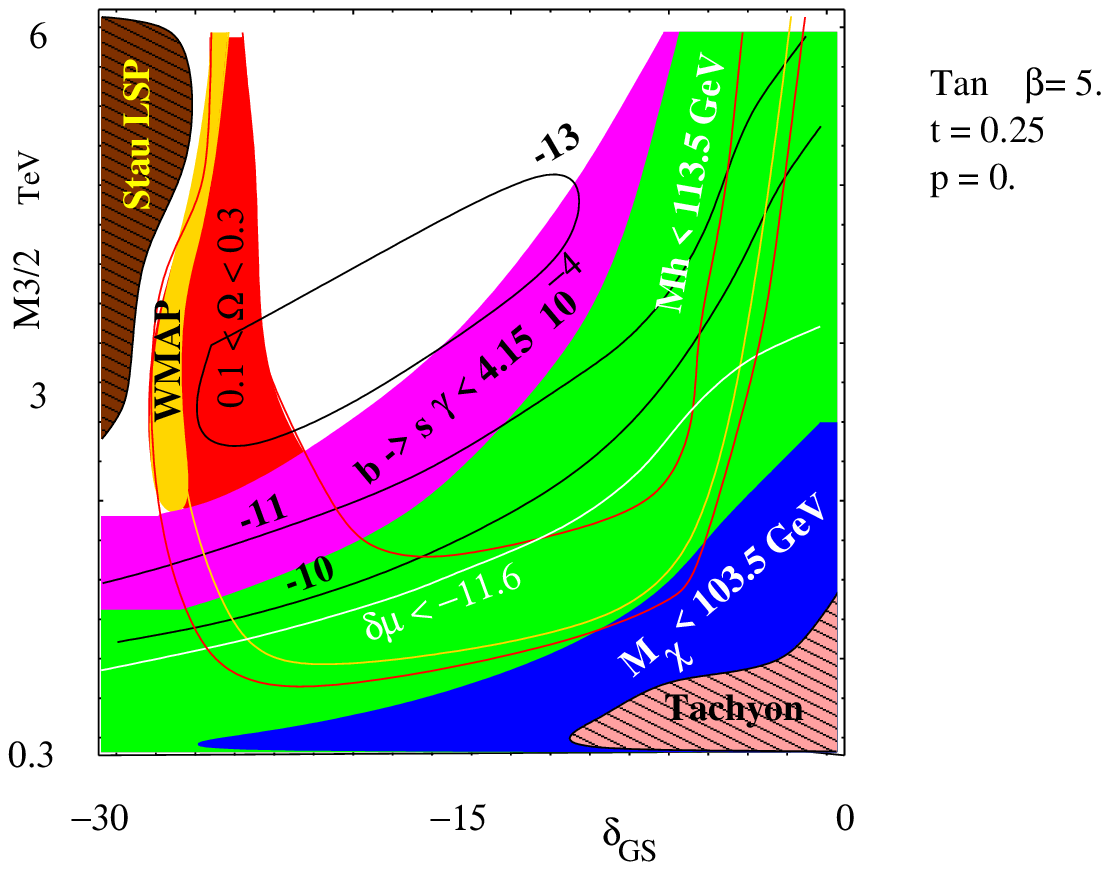,width=0.45\textwidth}
       \epsfig{file=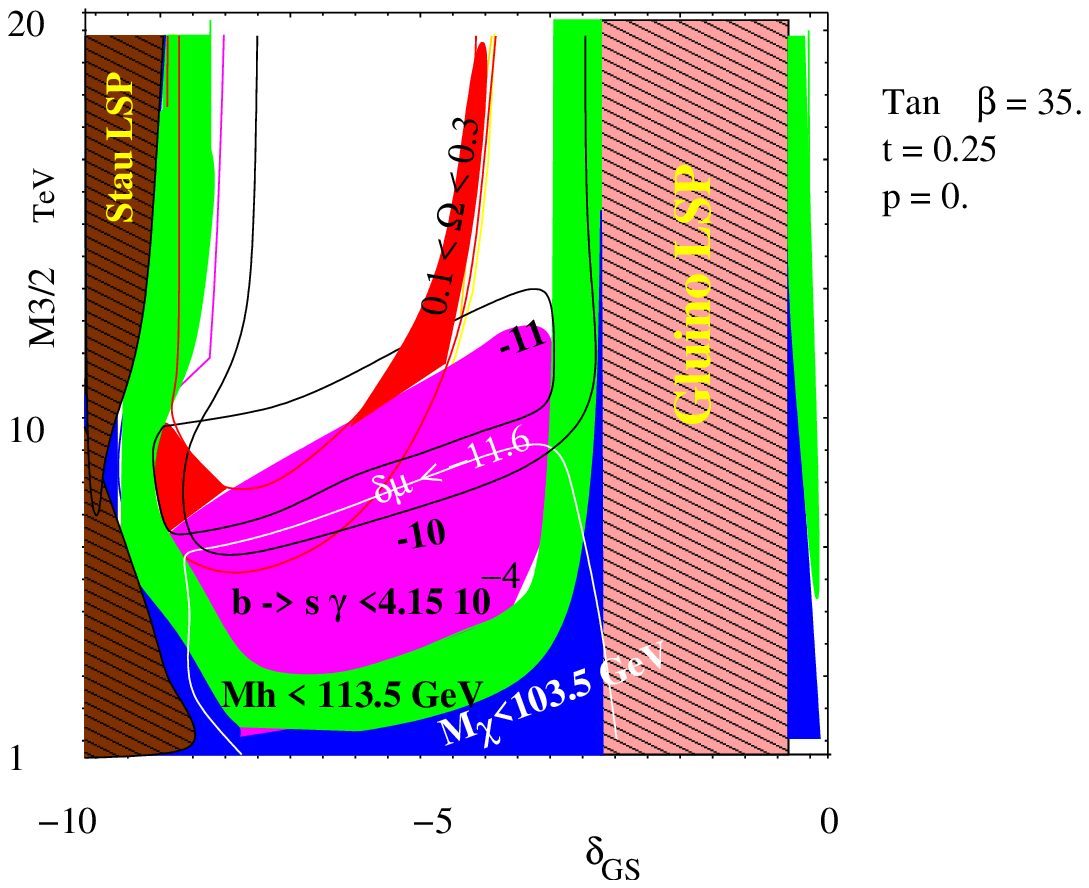,width=0.43\textwidth}}
          \caption{{\footnotesize {\bf The spin--independent scalar
          cross section in the moduli
          parameter space in the ($\delta_{GS}$, $M_{3/2}$) plane, 
          for $t=0.25,~p=0, ~\mu > 0$}, for tan $\beta$ $=5$ (left) 
          and $35$ (right).
          The labels on the black lines correspond to the $Log_{10}$ value
          of $\sigma^{scal}_{\chi-p}$ (pb).
          For a description of the experimental constraints applied, see  
          Section~\ref{sec:constraints}.}}
        \label{fig:dirmoduli_dGS}
    \end{center}
\end{figure}

\begin{figure}
    \begin{center}
\centerline{
       \epsfig{file=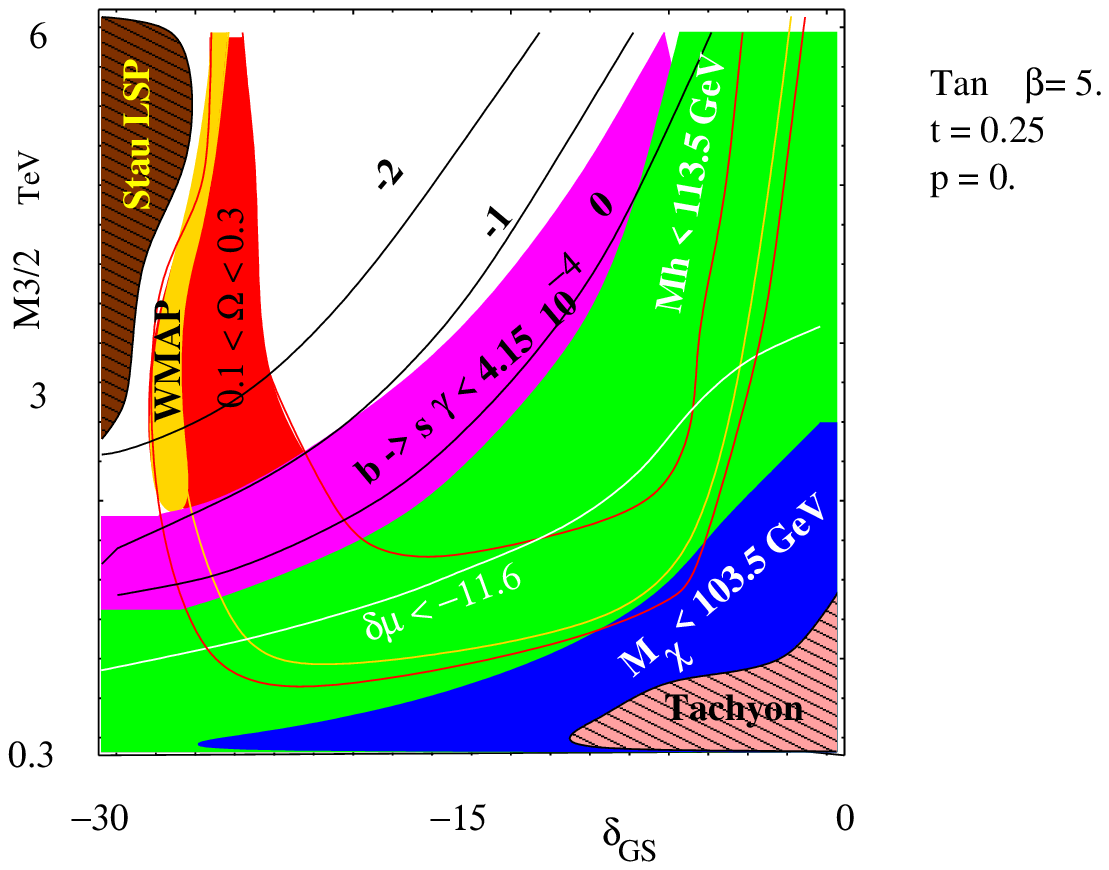,width=0.45\textwidth}
       \epsfig{file=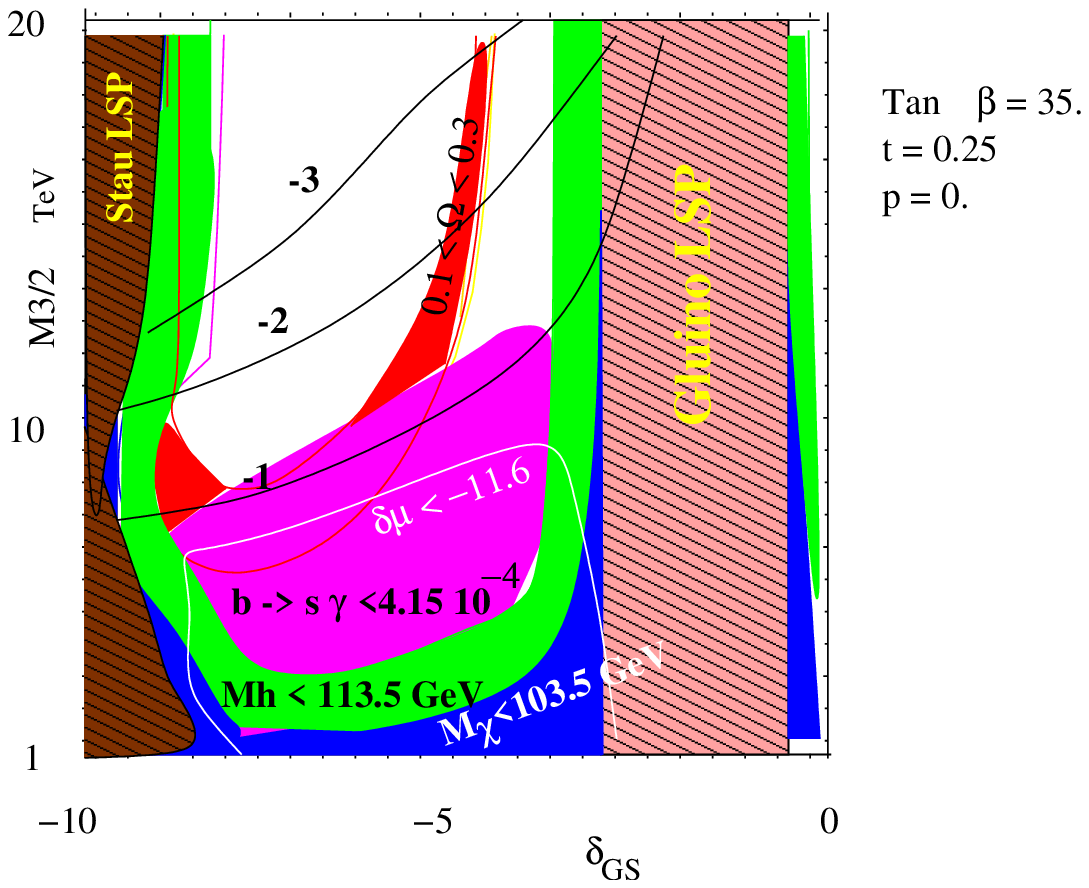,width=0.43\textwidth}}
          \caption{{\footnotesize {\bf Muon fluxes from the sun in the moduli
          parameter space in the ($\delta_{GS}$, $M_{3/2}$) plane, 
          for $t=0.25,~p=0, ~\mu > 0$}, for tan $\beta$ $=5$ (left) 
          and $35$ (right). 
          The labels on the black lines correspond to the $Log_{10}$ value
          of the flux ($ \mu\ {\rm km^{-2}\ yr^{-1}}$).
          For a description of the experimental constraints applied, see  
          Section~\ref{sec:constraints}.}}
        \label{fig:indmoduli_dGS}
    \end{center}
\end{figure}

We present in Figs \ref{fig:dirmoduli_p}--\ref{fig:indmoduli_dGS} the 
parameter space allowed by experimental constraints and the corresponding 
values of two important observables in the context of dark matter : the
spin-independent scalar cross-section $\sigma^{scal}_{\chi-p}$ and the muon 
flux from the Sun. 

Let us start with a general comment concerning the relic density. As we 
can see looking back at (\ref{modsoftgaugi}) and (\ref{modsoftscal}), 
the main feature of the moduli-dominated regime is to have gaugino $and$ 
scalar masses mediated by one loop corrections and threshold effects. 
This implies a phenomenology with light neutralinos and possibly light squarks.
In any case, it has been shown in \cite{Bin1} that the lightest neutralino 
$\chi$ keeps a wino nature in a broad region of parameter space, being 
degenerated with the lightest chargino $ \chi^+_1$. The first effect of
this degeneracy is a complete depletion of the relic density which barely 
reaches $10^{-2}$ due to the strong coannihilations channel ($\chi\chi^+_1$). 
The only way of splitting these two masses is  by the influence of the 
universal negative Green--Schwarz counterterm $\delta_{GS}$ in 
(\ref{modsoftgaugi}). Increasing $|\delta_{GS}|$  decreases the ratio 
$M_1/M_2$ leading first to the critical value $M_1|_{l.e.}/M_2|_{l.e.}=1$ 
around which the bino and wino contents give an interesting relic density. 
By increasing further $|\delta_{GS}|$, we can reach a bino--like region, where 
the relic density is enhanced up to the point where the neutralino mass
reaches the lightest stau mass giving a density  compatible with the
last WMAP results ($\tilde{\tau}$ coannihilation corridor). This is very 
well illustrated on the regions denoted $0.1<\Omega<0.3$ and WMAP in Figs. 
\ref{fig:dirmoduli_dGS} and \ref{fig:indmoduli_dGS}. 


Figures \ref{fig:dirmoduli_p} and \ref{fig:indmoduli_p} (left) present 
the space constraints in a  ($t,M_{3/2}$) plane for $\delta_{GS}=0$, 
$p=0$ and $\tan{\beta}=35$. With this choice of parameters, the lightest 
neutralino is always wino-like, independently of the value of $t\equiv\langle
T \rangle$. Indeed, 
the gaugino masses $M_a$ in \ref{modsoftgaugi} are proportional to 
their respective beta--function coefficients $b_a$, which yields the 
standard relation, at the GUT scale :
\begin{equation}
\frac{M_1}{M_2} \sim \frac{b_1}{b_2} = \frac{33}{5} =6.6
\end{equation}
implying $M_2 < M_1$ at the electroweak scale. This leaves the lightest
neutralino $\chi$ and chargino $ \chi^{\pm}_1$ in a wino state, which
depletes the relic density of the neutralino ($10^{-2}$ at its maximum level 
value). 

If we decrease the value of $t$ from its self--dual point $t=1$, the function 
$G_2(t,\overline{t})$ will reach a point where $M_{i=1,2}=0$, corresponding to
\begin{equation}
(t + \overline{t}) G_2(t,\overline{t})=1 \rightarrow \mathrm{Re}~t=0.523
\end{equation}
giving a null value for gaugino mass terms, and, as a consequence,
for the scalar neutralino-proton cross section (proportional to $m_{\chi}$). 
as shown in Fig. \ref{fig:dirmoduli_p}. After this pole, 
$M_{i=1,2}$ becomes negative, and its absolute value increases with $t$ : this
restores a non-vanishing cross-section. This cross-section 
falls with $t$ becoming small because all the sparticles (even the scalars
driven by $M_3$ through the renormalization group equations) become more
massive ($G_2 \rightarrow \infty$ as $t \rightarrow 0$), which increases 
the virtuality of the exchanged squarks or neutral Higgs in the elastic 
diffusion process.  
Generically, for a wino-like neutralino, neutrino flux coming from the
Sun are small although annihilations into gauge bosons (which give more
energetic neutrinos than other annihilation channels) are favoured. 
Indeed the capture is small due to suppression of the $Z$ exchange, 
proportional to the higgsino content of the neutralino. This explains 
the small values of the fluxes in the $(t,M_{3/2})$ plane. 
This is illustrated on the left panel of Fig. \ref{fig:indmoduli_p}
where muon fluxes (which we will denote by ${\rm flux}^{\odot}_{\mu}$) 
coming from the Sun are smaller than 
$ {\rm 10\ km^{-2}\ yr^{-1}}$, well below 
possible detection.



We show on the right panel of Figs. \ref{fig:dirmoduli_p} and 
\ref{fig:indmoduli_p} the effects of the parameter $p$ and $\tan{\beta}$ 
for a fixed value of the gravitino mass ($M_{3/2}=40$ TeV) keeping 
$\delta_{GS}=0$. From (\ref{modsoftscal}), we see that increasing $p$ up 
to one decreases scalar soft masses, increasing the scalar cross section 
$\sigma^{scal}_{\chi - p}$ : for p=0.95, the scalars are sufficiently light 
(\ref{modsoftscal}) to allow a cross section of the order of $10^{-8}$ pb 
in some region of the parameter space. 
For high values of  $\tan{\beta}$, $m_H$ decreases and one has lower values
 of $\mu$ so higher higgsino content of the neutralino before the ``noEWSB'' 
boundary which also enhances the Higgs coupling in $\sigma^{scal}_{\chi - p}$ 
(proportional to $z_{\chi,i=1,2} z_{\chi,i=3,4}$). Along this boundary, 
although it is excluded by limits on $b\rightarrow s\gamma$ and $m_h$, fluxes 
coming from the Sun for neutrino indirect detection can be high. 
For small $\tan{\beta}$, the $\sigma^{scal}_{\chi - p}$ enhancement comes 
from the light Higgs contribution $\chi q \xrightarrow{h} \chi q$. 

Whereas direct detection, being driven by Higgs ($h,H$) exchange has a
strong $\tan{\beta}$ dependance as can be seen on Fig. \ref{fig:dirmoduli_p},
the indirect detection  ($Z$ exchange in the capture) is clearly independent 
of  $\tan{\beta}$ as shown in Fig. \ref{fig:indmoduli_p}.


\begin{figure}
    \begin{center}
\centerline{       \epsfig{file=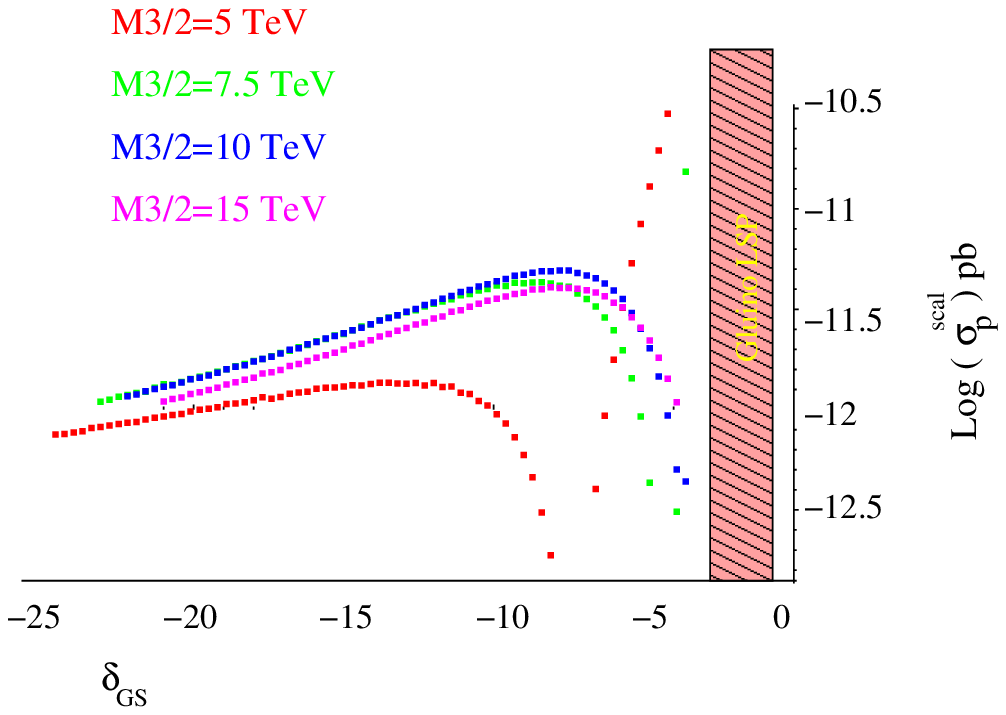,width=0.45\textwidth}
       \epsfig{file=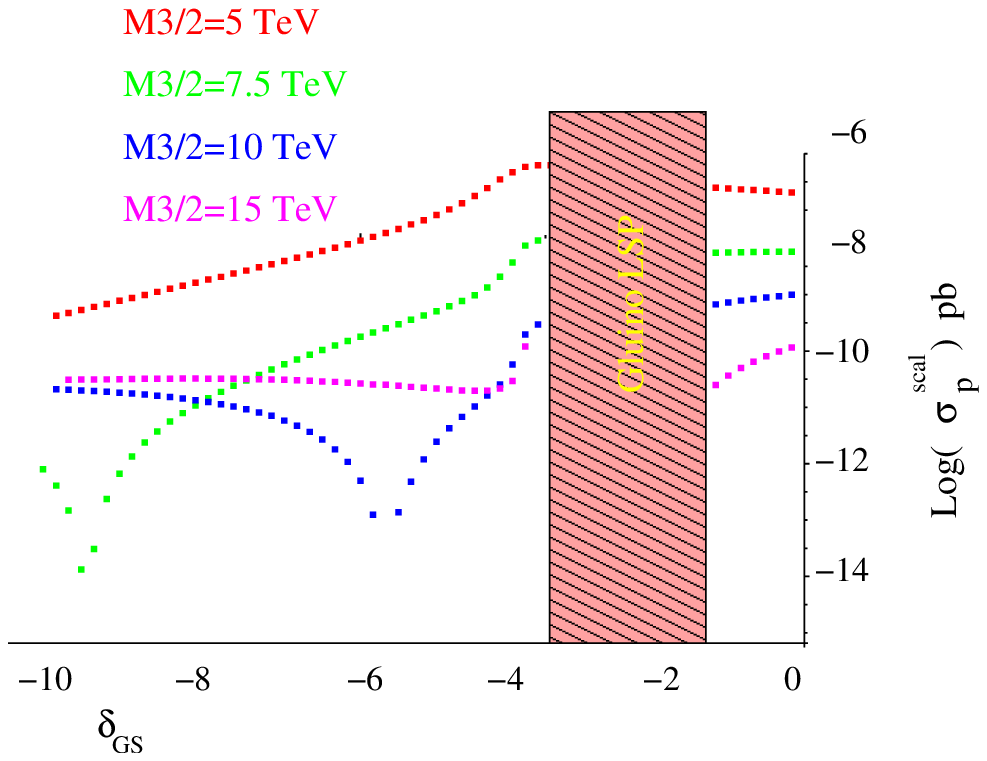,width=0.45\textwidth}}
          \caption{{\footnotesize {\bf The spin--independent scalar
          cross section as a function of the 
          Green-Schwarz counterterm $\delta_{GS}$}, for $t=0.25$, $p=0$
           and different values 
          of $M_{3/2}$, with tan $\beta=5$ (left) and 35 (right).}}
          \label{fig:Crossdgs}
    \end{center}
\end{figure}

\begin{figure}
    \begin{center}
\centerline{
       \epsfig{file=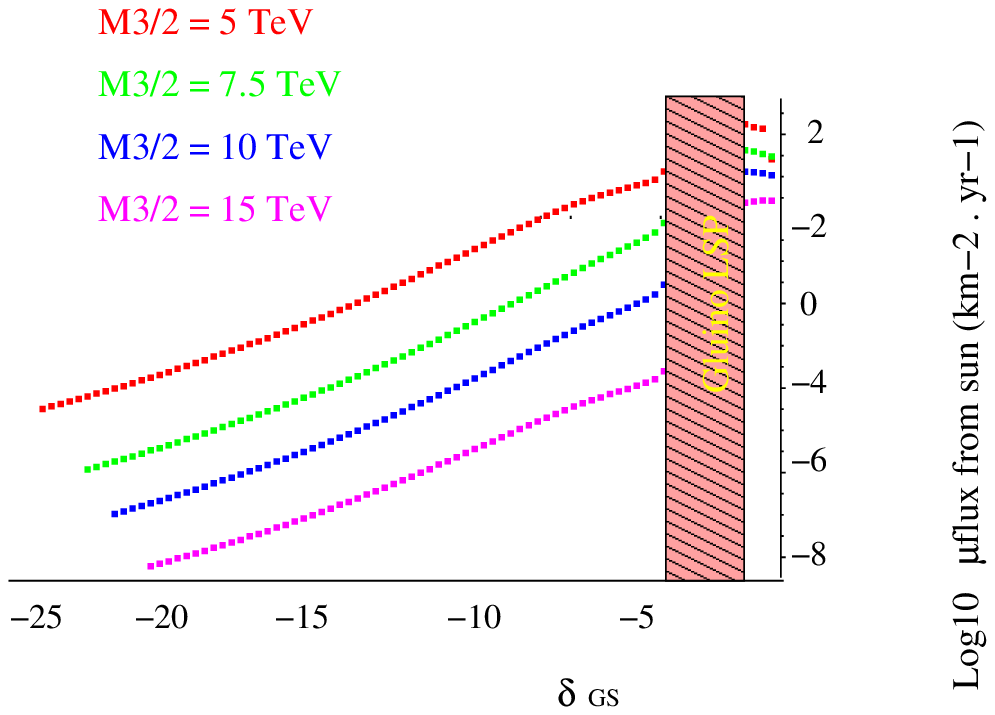,width=0.45\textwidth}
       \epsfig{file=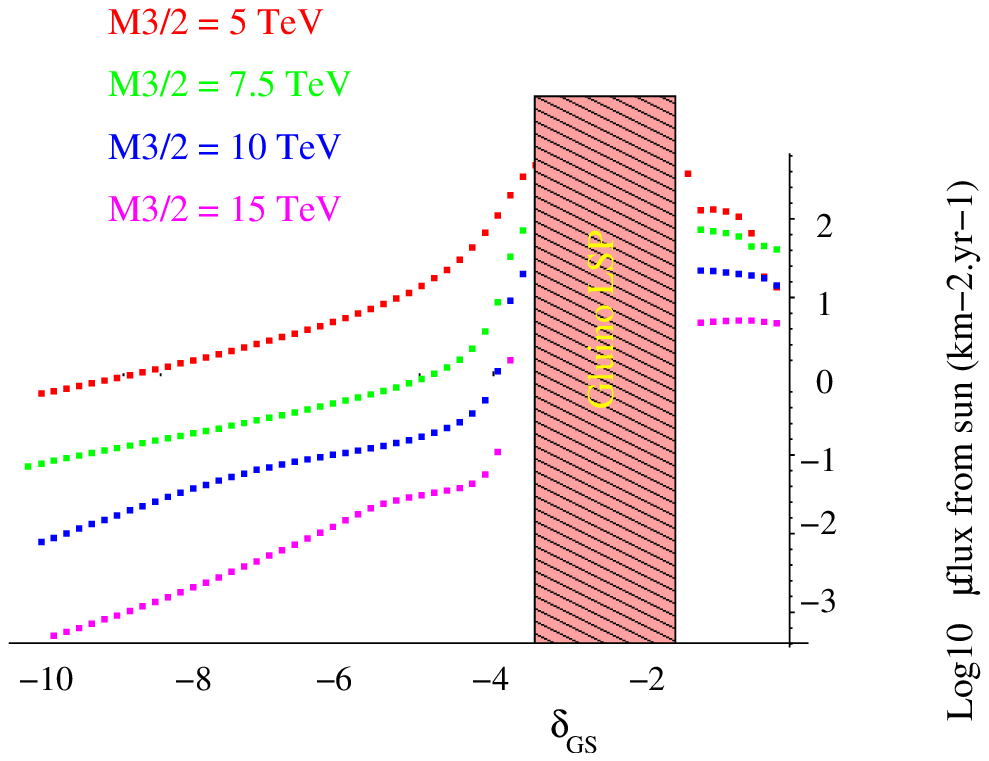,width=0.45\textwidth}}
          \caption{{\footnotesize {\bf Muon fluxes from the sun 
          as a function of the 
          Green-Schwarz counterterm $\delta_{GS}$}, for $t=0.25$, $p=0$
           and different values 
          of $M_{3/2}$, with tan $\beta=5$ (left) and 35 (right).}}
          \label{fig:Fluxdgs}
    \end{center}
\end{figure}

Figures \ref{fig:dirmoduli_dGS} and  \ref{fig:indmoduli_dGS} show,
respectively for direct detection and indirect detection, the
($\delta_{GS},M_{3/2}$) plane for $t=0.25$, $p=0$ and $\tan{\beta}=5$ or 35. 
Because of the positive sign of $b_1$, the effect of $\delta_{GS}$ 
(which is negative) allows the neutralino to be bino-like and splits
$m_{\chi}$ and $m_{ \chi^+_1}$. Indeed, looking back at (\ref{modsoftgaugi}), 
we can see that, at grand unification,  
$\frac{M_1}{M_2}=\frac{\alpha b_1 -\delta_{GS}}{\alpha b_2 -\delta_{GS}}$,
with $\alpha$ independent of the gauge group considered. 
Decreasing from zero $\delta_{GS}$ decreases the wino (versus the bino) 
content of the neutralino and thus increases the relic density. For 
$\delta_{GS} \sim -23/5$,  we have the relation 
$M_1|_{l.e.}/M_2|_{l.e.} \sim 1$ at low 
energy giving $\Omega h^2|_{\chi} \sim \Omega h^2|^{WMAP}_{CDM}$. Taking larger
$|\delta_{GS}|$ values means driving to the universal case, with 
$M_1|_{{\scriptsize GUT}}=M_2|_{{\scriptsize GUT}}$ at the
high energy scale, and so $M_1|_{l.e.} \sim 0.6 M_2|_{l.e.}$ at the 
electroweak scale, giving back a bino-like lightest neutralino.
Cosmologically favoured relic densities then come from $\chi \tilde{\tau}$ 
coannihilation along the $\tilde{\tau}$ LSP boundary line. However in this
region, fluxes are very small as capture is disfavoured by squark
(though quite light) exchange and very small higgsino content of neutralino in
Higgs exchange.

Decreasing $\delta_{GS}$ first decreases $|M_3|$ down to gluino LSP points and
then increases $|M_3|$ (which is now negative), scalar masses (through $M_3$ 
RGE effect) and $\mu$ (decreasing the neutralino higgsino fraction). 
In particular the lightest Higgs $h$ (through its
stop radiative correction) and the heavy neutral Higgs $H$ masses increase
 and the coupling $z_{\chi,1(2)}z_{\chi,3(4)}$ decreases so that
$\sigma^{scal}_{\chi - p}$ decreases following the higgs mass experimental
limit contour (see Fig. \ref{fig:dirmoduli_dGS}). For high values of $M_{3/2}$,
$ \sigma^{scal}_{\chi - p}$ is actually  beyond reach of experiment 
sensitivities ($\leq 10^{-11}$ pb).
Moreover, for small values of 
$\delta_{GS}$, $\sigma^{scal}_{\chi -p}$  follows the contour of iso--$\mu$ 
values (the higgsino content of neutralino) along the
$m_{\chi^+_1}$ limit. Higher values of $\tan{\beta}$ are also more
favourable to direct detection because the Higgs $H$ is lighter.
For low values of $M_{3/2}$, $\mu$ values are smaller and the higgsino 
fraction of 
neutralino drives the phenomenology. Relic density is closed to the WMAP 
range and both direct ($\sigma^{scal}_{\chi - p}\sim10^{-8}$ pb in 
Fig. \ref{fig:dirmoduli_dGS}) and neutrino indirect detection 
(${\rm flux}^{\odot}_{\mu}\sim10^{3-2}\ {\rm km^{-2}yr^{-1}}$ in  
Fig. \ref{fig:indmoduli_dGS}) can be interesting but this region
is excluded by limits on $m_h$ and $b\rightarrow s\gamma$. Muon fluxes coming 
from the Sun follow the iso--$\mu$ shape given by the limit on $m_{\chi_1^+}$. 
Small $|\delta_{GS}|$ can also lead to gluino LSP.

\begin{figure}[t!]
    \begin{center}
\centerline{
       \epsfig{file=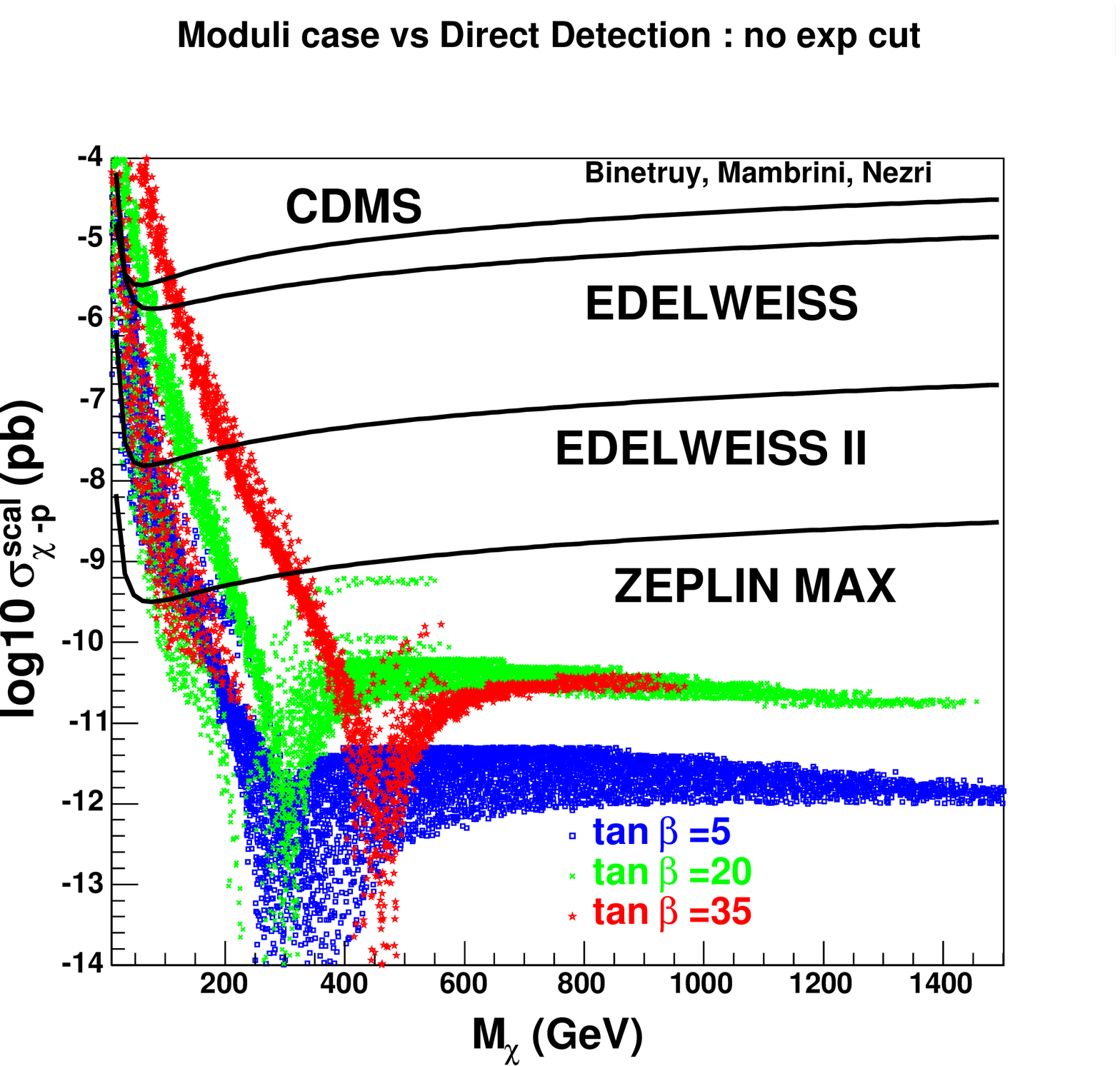,width=0.45\textwidth}
       \epsfig{file=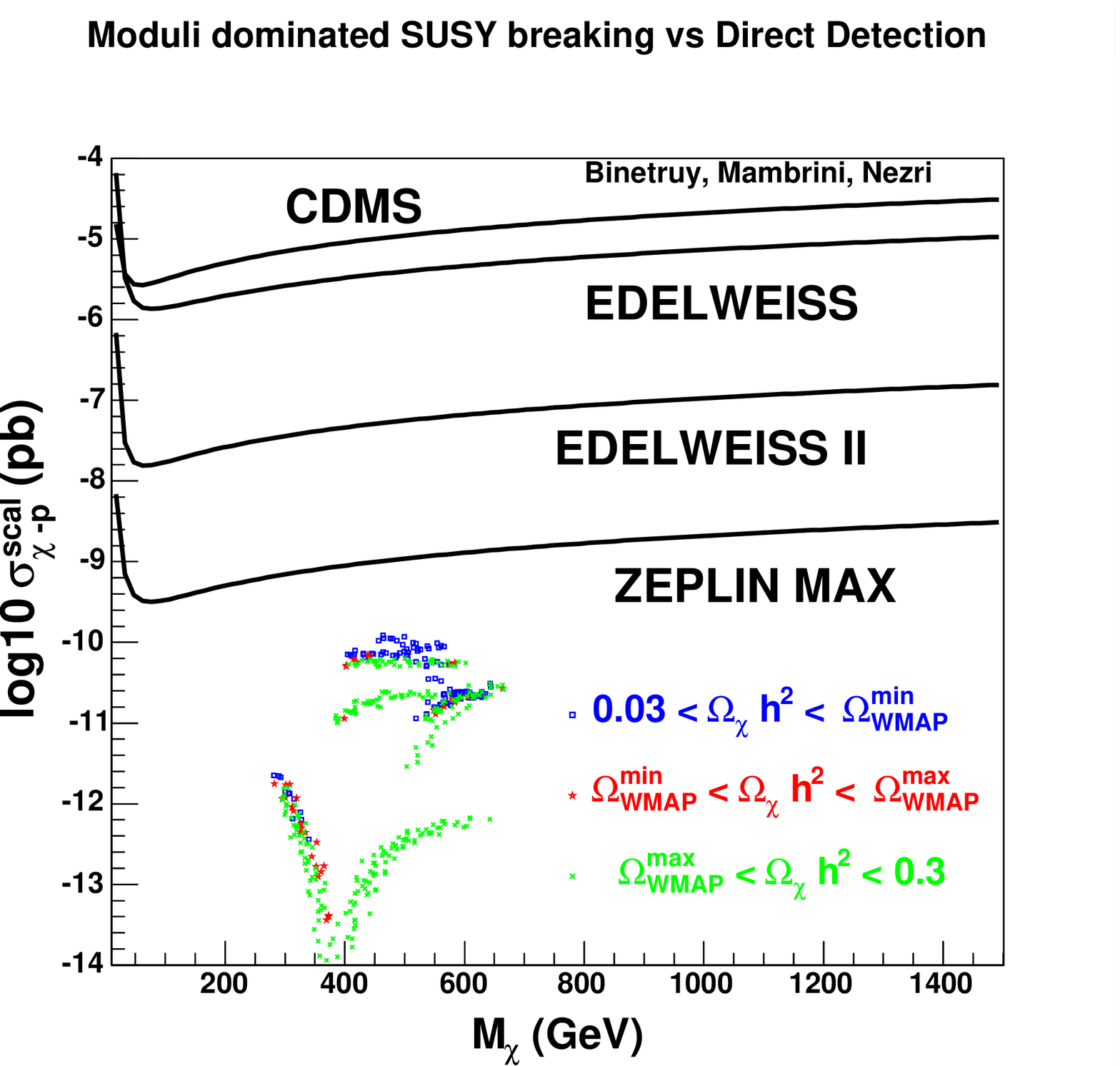,width=0.45\textwidth}}
          \caption{{\footnotesize {\bf The spin--independent scalar
          cross section as a function of the neutralino mass $M_{\chi}$}
          for $t=0.25$, $p=0$ and
          $\tan\beta=5,~20,~35$. We have scanned the moduli parameter space
          on $M_{3/2}$ and $\delta_{GS}$ before (left) and after (right) 
          having applied the accelerator and cosmological constraints of
           Section~\ref{sec:constraints}.}}
          \label{fig:Sigma_mchi}
    \end{center}
\end{figure}


We have illustrated explicitly the $\delta_{GS}$ dependance of
$\sigma^{scal}_{\chi-p}$ and ${\rm flux}^{\odot}_{\mu}$ on Figs. 
\ref{fig:Crossdgs} and \ref{fig:Fluxdgs}. As previously said, decreasing 
$\delta_{GS}$ increases 
scalar masses (through $M_3$ running effect) so $m_H$ which dominantly drive
$\sigma^{scal}_{\chi-p}$. It also exchanges wino and bino content leading 
to smaller coupling ($\tan{\theta_W}$ suppressed) and lower values of 
$\sigma^{scal}_{\chi-p}$. We also see the well known strong influence
of $\tan{\beta}$ on $m_H(m_A)$ leading to higher values for
$\tan{\beta}=35$. 

Concerning indirect detection, bino content decreases the fluxes but small 
values of $|\delta_{GS}|$ can lead to gluino LSP and high fluxes nearby. 
This corresponds to the  effect of small values of $M_3|_{GUT}$ in
soft parameter running, leading to small values of $\mu$ and 
favoring both direct and indirect detection \cite{Mynonuniv}. 
But in the models we consider in this
section, this happens in region excluded by accelerator constraints
because of light chargino contribution in $b \rightarrow s \gamma$ and too 
light Higgs $h$.

\begin{figure}[t!]
    \begin{center}
 \centerline{
\epsfig{file=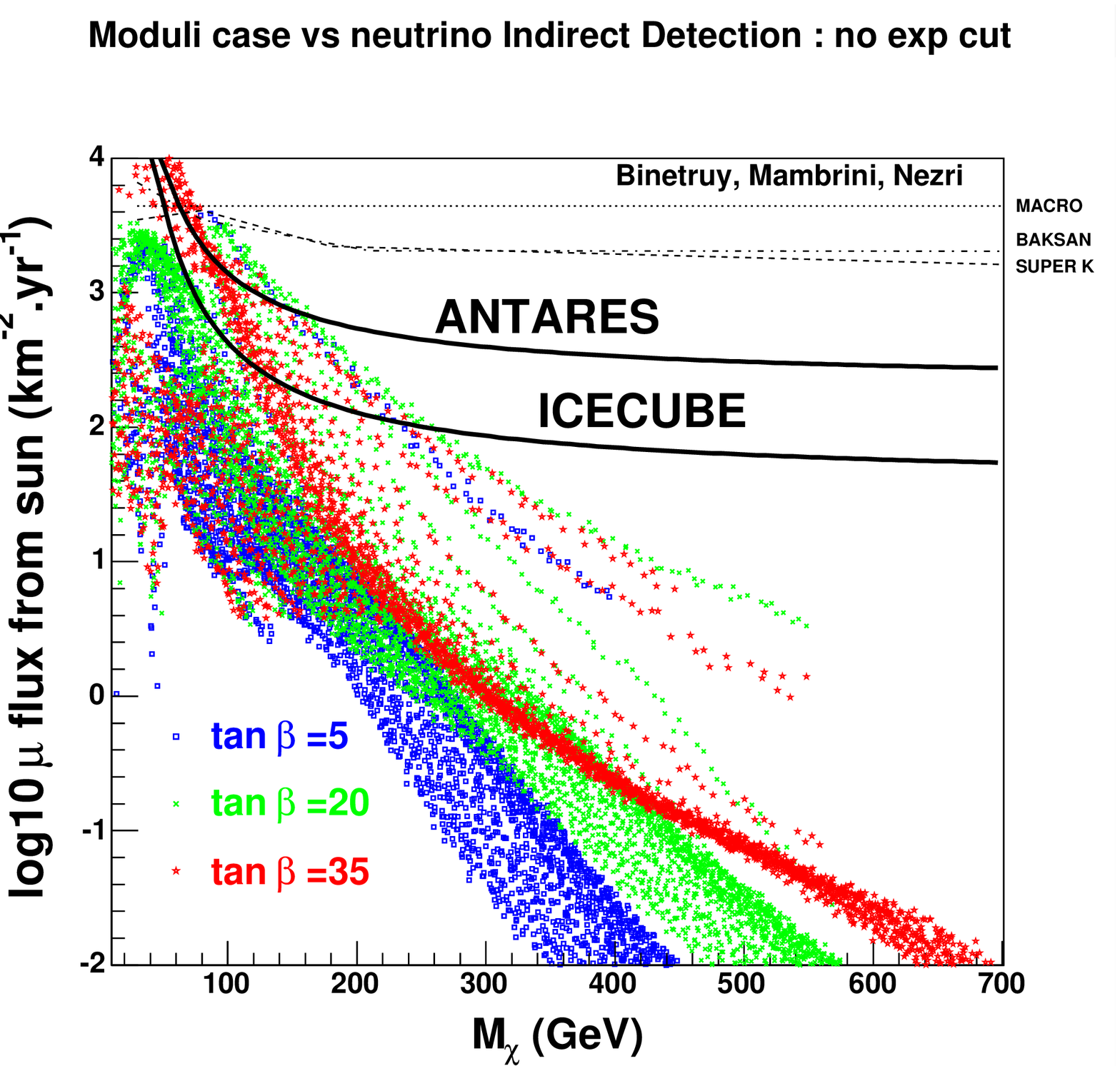,width=0.45\textwidth}
\epsfig{file=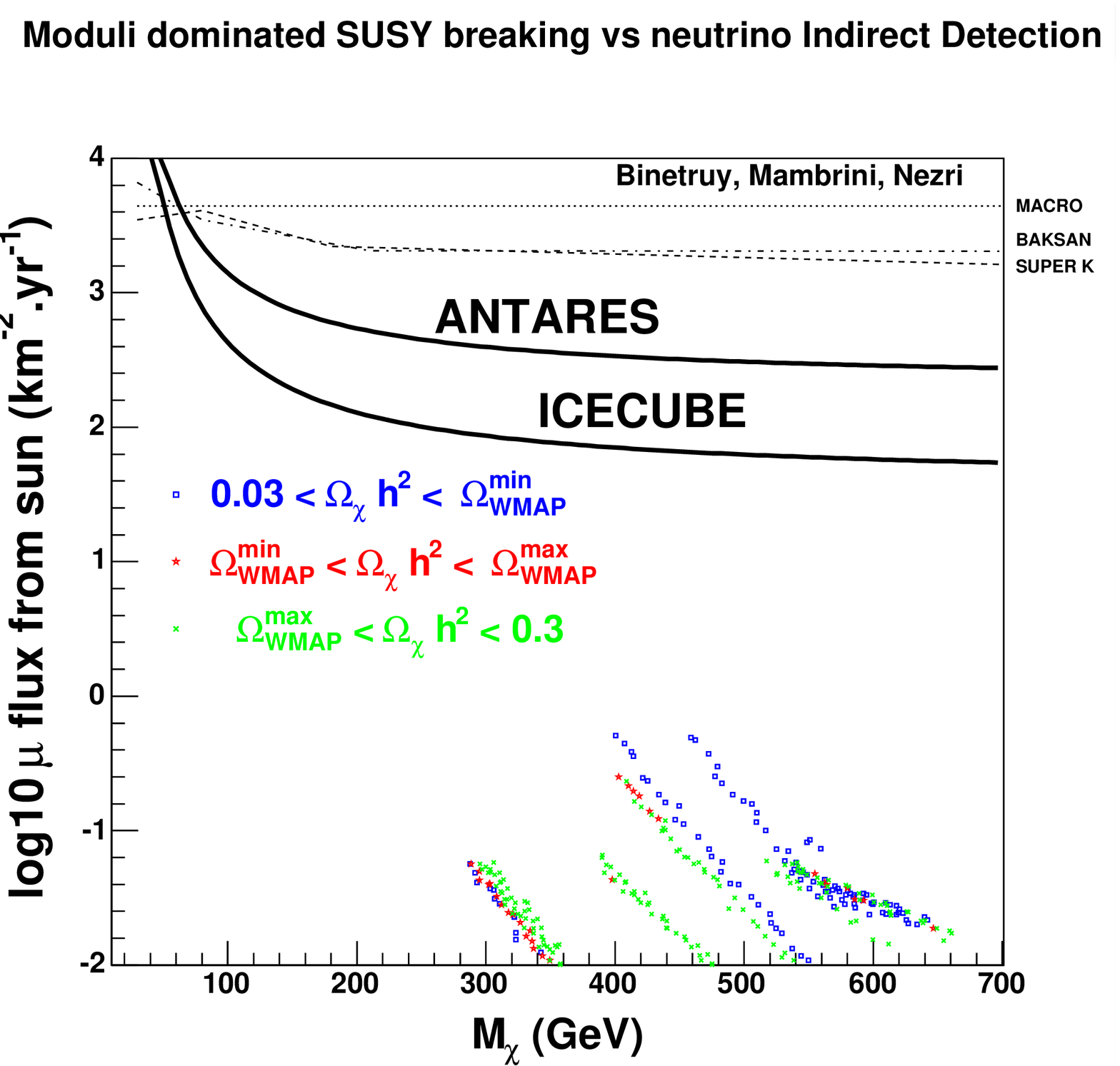,width=0.45\textwidth}}
          \caption{{\footnotesize {\bf Muon fluxes from the sun
          as a function of the neutralino mass $M_{\chi}$}
          for $t=0.25$, $p=0$ and
          $\tan\beta=5,~20,~35$. We have scanned the moduli parameter space
          on $M_{3/2}$ and $\delta_{GS}$ before (left) and after (right) 
          having applied the accelerator and cosmological constraints of
           Section~\ref{sec:constraints}.}}
          \label{fig:Flux_mchi}
    \end{center}
\end{figure}
\vskip .5cm
To conclude the discussion of the moduli-dominated SUSY breaking case, 
we show a wide range of models in the $(m_{\chi},\sigma^{scal}_{\chi-p})$ 
plane for direct detection (Fig. \ref{fig:Sigma_mchi}) and in the
$(m_{\chi},{\rm flux}^{\odot}_{\mu})$ plane for neutrino indirect detection
(Fig. \ref{fig:Flux_mchi}). We then clearly see the complementarity
between dark matter and accelerator searches by comparing left panel
(no accelerator constraints) and right panel (accelerator constraints
applied) of these two figures.

In regions satisfying accelerator constraints, moduli dominated
model give rise to either wino neutralino with strongly suppressed
relic density  and very low detection rates, or bino neutralino (thanks
to $\delta_{GS}$) and interesting relic density via 
$\chi \tilde{\tau}$ coannihilation or adjusted wino vs bino content but 
again the direct and indirect detection rates are beyond reach of detection 
(except sometimes for a larger than one ton size direct detection experiment 
such as Zeplin Max). This situation is summarized on Figs. 
\ref{fig:Sigma_mchi} and \ref{fig:Flux_mchi}. 

One interesting feature of this scenario is that $M_{i=1,2}$ is
becoming negative for large values of $|\delta_{GS}|$. This negative
sign can allow cancellation between the up--quark and down--quark
contribution to the scalar cross section
$\sigma_{\chi-p}^{\mathrm{scal}}$ together as for negative values of
$\mu$ pointed in \cite{Ellismuneg}.
This phenomenon is responsible of the "seagull" shape
of the plots \ref{fig:Crossdgs} and \ref{fig:Sigma_mchi}, depleting
in the same way the scalar cross section down to $10^{-13}$ (for
tan$\beta$=5) or $10^{-11}$ (for tan$\beta$=35).

Moduli models satisfying accelerator constraints and a WMAP favoured relic 
density  are generically not detectable by dark matter searches.

\subsection{Dilaton dominated case}

The phenomenology of the dilaton dominated scenario is completely different
from the moduli domination just discussed. If we look at Eqs 
(\ref{dilatsoftscal}) and (\ref{dilatsoftgaugi}), it
is clear that we are in a domain of heavy squarks and sleptons (of the order
of the gravitino scale) and light gaugino masses (determined by the dilaton
auxiliary component $vev$). Indeed, the beta--functions $b_a$ being of 
the order of $10^{-2}$, the corresponding terms are not competitive by 
comparison to the $F$ term of the dilaton in (\ref{dilatsoftgaugi}). 
In fact, looking more closely at (\ref{Ktrue}) and
(\ref{FS}), for not too large values of the universal beta--function 
coefficient of the first condensing group  ($b_+$), we may consider that 
$F^S$ is a linear function of $b_+$. Increasing $b_+$ means approaching
 the universal case for the gaugino sector (and the scalar one, driven by 
 $M_{3/2}$).

\begin{figure}
    \begin{center}
\centerline{
       \epsfig{file=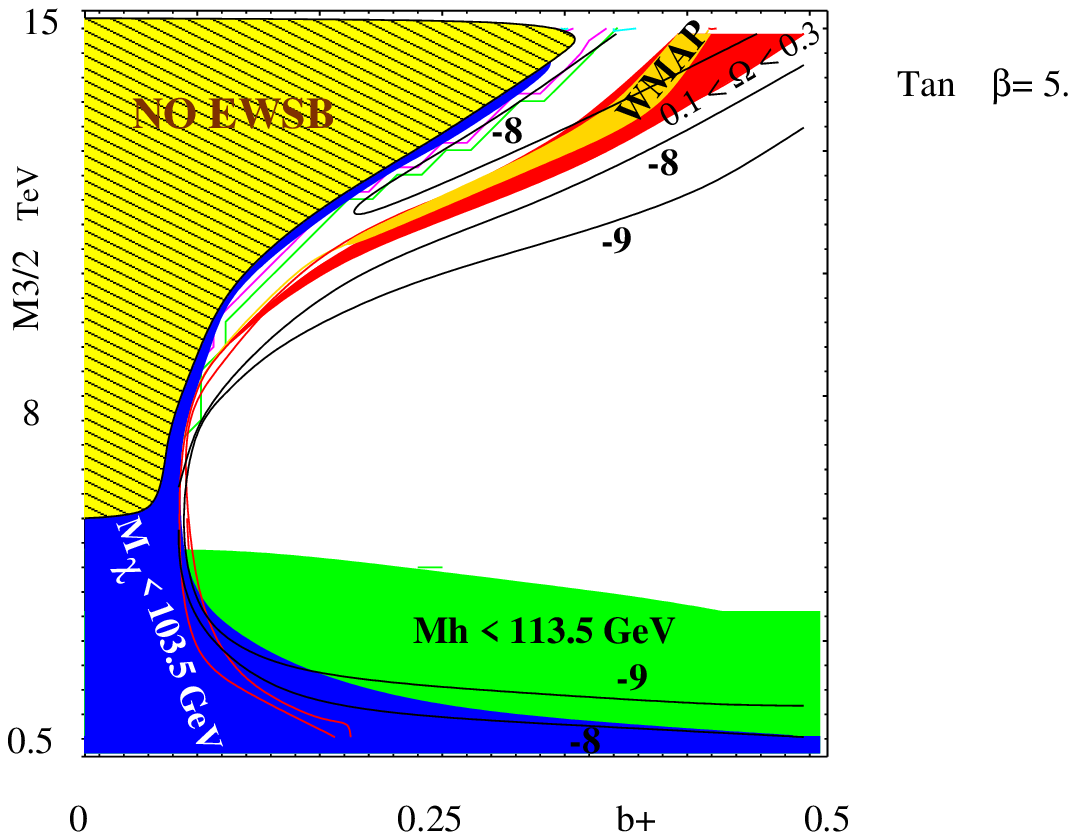,width=0.45\textwidth}
       \epsfig{file=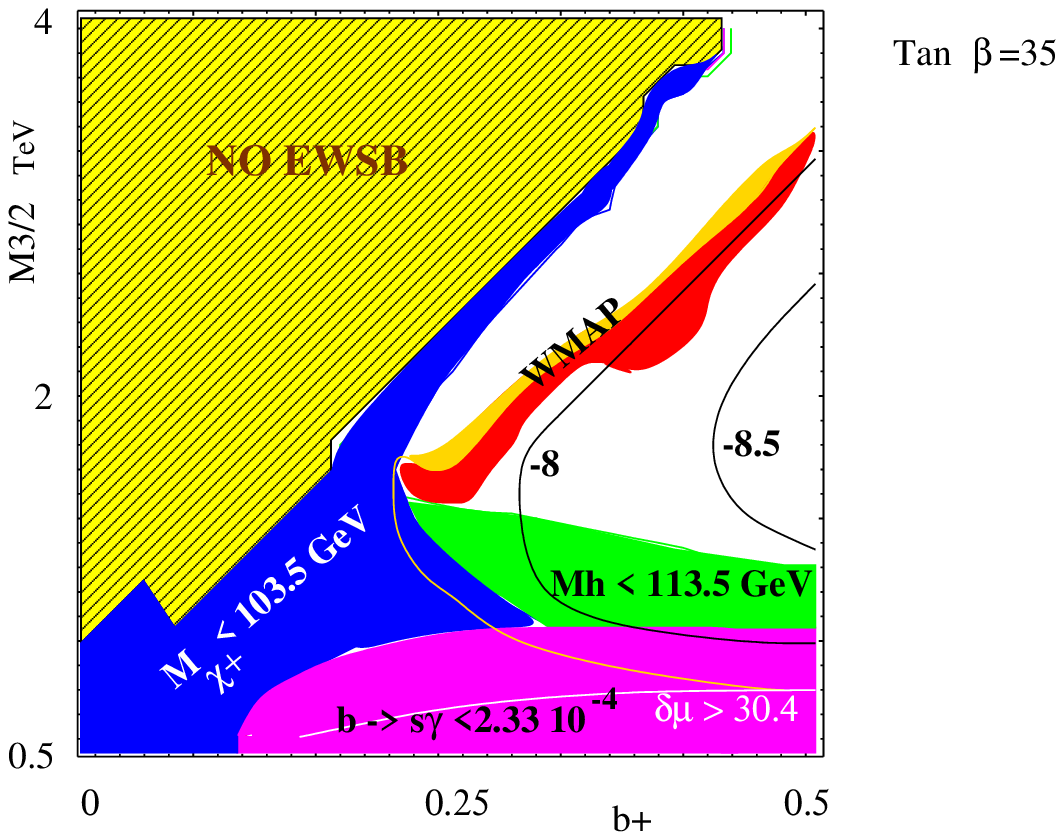,width=0.45\textwidth}}
          \caption{{\footnotesize {\bf The spin--independent scalar
          cross section in the dilaton
          parameter space}, in the ($b_+$, $M_{3/2}$)
          plane, for tan $\beta$ $=5$ (left) and tan $\beta$ $=35$ (right).
          Accelerators and
          cosmological constraints are given for $\mu > 0$. 
          The labels in the black lines correspond to the $Log_{10}$ value
          of $\sigma^{scal}_{\chi-p}$ (pb).
          For a description of
          the experimental constraints applied, see 
          Section~\ref{sec:constraints}.}}
          \label{fig:dirdilatonb+}
    \end{center}
\end{figure}

\begin{figure}
    \begin{center}
\centerline{
       \epsfig{file=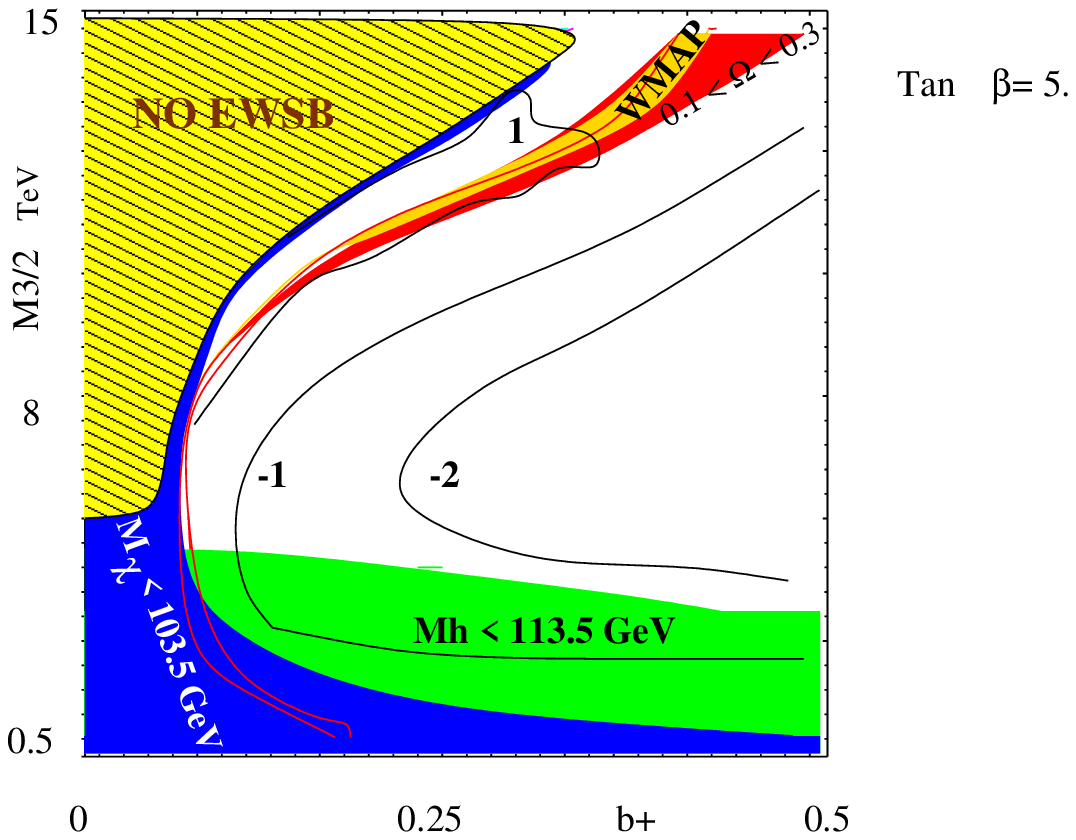,width=0.45\textwidth}
       \epsfig{file=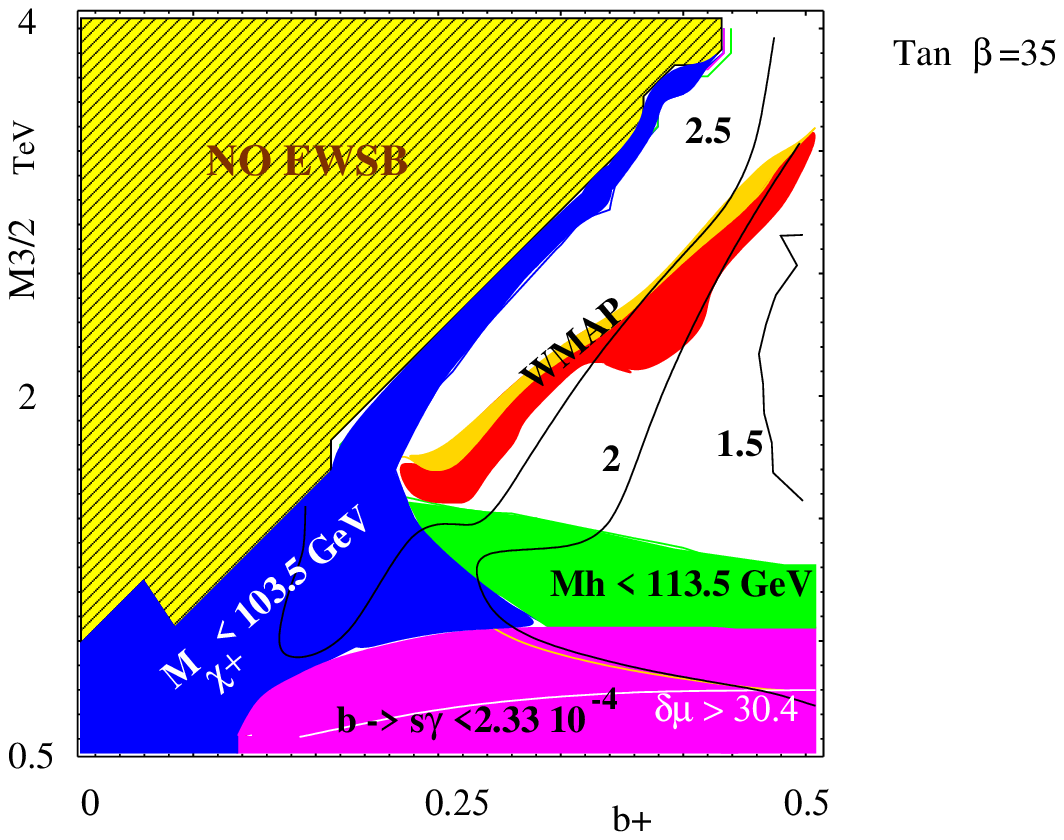,width=0.45\textwidth}}
          \caption{{\footnotesize {\bf Muon fluxes from the sun 
          in the dilaton
          parameter space in the ($b_+$, $M_{3/2}$) plane} for
          $\tan\beta=5$ (left) and $35$ (right).Accelerators and
          cosmological constraints are given for $\mu > 0$.  The
          labels in the black lines correspond to the $Log_{10}$ value 
        of the flux ($ \mu\ {\rm km^{-2}\ yr^{-1}}$).
          For a description of the experimental constraints applied, see 
          Section~\ref{sec:constraints}.}}
          \label{fig:Fluxdilaton}
    \end{center}
\end{figure}

Figures \ref{fig:dirdilatonb+} and \ref{fig:Fluxdilaton} present the
$(b_+,M_{3/2})$ plane for $\tan{\beta}=5$ and 35 with experimental
exclusions, neutralino relic density and respectively 
iso-$\sigma^{scal}_{\chi-p}$ and iso-${\rm flux}^{\odot}_{\mu}$
curves. We first discuss the experimental exclusion plots.

\begin{figure}
    \begin{center}
\centerline{
       \epsfig{file=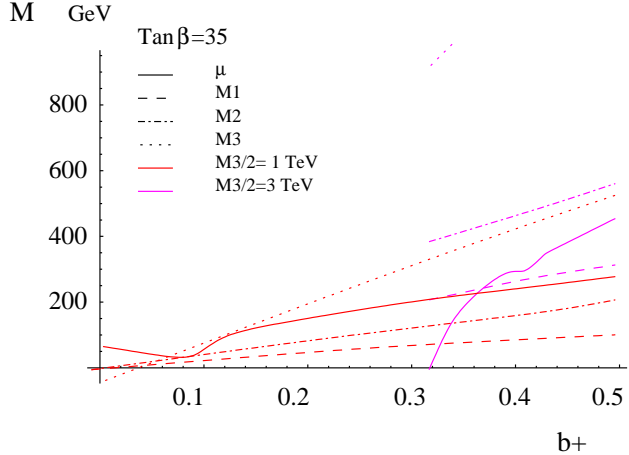,width=0.5\textwidth}}
          \caption{{\footnotesize  Low energy parameter $M_1$, $M_2$, $M_3$, 
          $\mu$ evolution 
          with $b_+$, for $\tan{\beta}=$ 35 and $M_{3/2}$=1000,
          3000\ {\rm GeV}.}}
          \label{fig:m1m2m3mudilat}
    \end{center}
\end{figure}

For low values of $M_{3/2}$, scalars and gauginos are light
so that accelerator constraints are strong. At fixed values of $M_{3/2}$
and decreasing values of $b_+$, we see from (\ref{dilatsoftscal}) that
gaugino masses decrease \footnote{See Fig. \ref{fig:m1m2m3mudilat}. For 
very low values of $b_+$ around 0.1 (independent on $M_{3/2}$ value), 
we have a very localised range with gluino LSP due to cancellation in $M_3$ 
coming from the cancellation between $b_3$ ($<0$) and $b_+$.}. 
$M_3|_{GUT}$ being smaller, the $m^2_{H_u}$ running slope is softer and yields 
positive $m^2_{H_u}$ at low energy leading to lower values of $\mu$. This
explains the region with too light a chargino (i.e. Higgsino) mass followed by
the ``no EWSB'' region as one goes along a decreasing $b_+$ direction. 

As one increases $M_{3/2}$ at fixed $b_+$, one decreases $M_3$ and the same 
focussing effect as discussed in the previous paragraph leads to a lower 
value of the SUSY parameter $\mu$. It is thus not surprising that the 
``no EWSB'' extends further to the right as $M_{3/2}$ increases. 

We now turn to a discussion of the direct and indirect (neutrino) rates.
For low values of the gravitino mass, these rates are favoured because of 
the light squark contributions in
$\sigma^{\mathrm{scal}}_{\chi-p}$  and in
$\sigma^{\mathrm{spin}}_{\chi-p}$ (enhancing capture 
and ${\rm flux}^{\odot}_{\mu}$). \footnote{$b_+$ acts in the same way
since we have seen that it is increasing with $\mu$. 
But these low $M_{3/2}$ regions are ruled out by accelerator constraints.}

As one increases $M_{3/2}$, one finds a region where the lightest neutralino 
is a mixed higgsino--gaugino state. It can  satisfy WMAP requirements
on relic density through $\chi \chi \rightarrow W^+W^-,ZZ$ or $t\bar{t}$ 
annihilation and 
$\chi  \chi^+_1$,$\chi  \chi^0_2$ coannihilation processes. 
This region is safe from limits on $m_h$ and $b\rightarrow s\gamma$
thanks to the high $M_{3/2}$ values. 
Direct detection is favoured because the  higgsino component enhances
coupling in $\chi q \xrightarrow{H}\chi q$ leading to
 $\sigma^{scal}_{\chi - p}\sim 10^{-7-8}$ pb.
In this mixed region, indirect detection is also powerful (Fig. \ref{fig:Fluxdilaton}).
Indeed, the higgsino component allows efficient capture via
$\chi q \xrightarrow{Z} \chi q$ and neutralino annihilations into gauge
bosons or $t\bar{t}$ lead to energetic neutrinos/muons. Muon fluxes coming
from the Sun are enhanced and can reach $10^1$ to $10^{3}{\rm km^{-2}\ 
yr^{-1}}$. 
In both cases (direct or indirect detection), the detection rates increase with
tan$\beta$ because the higgsino fraction is higher, direct detection being
also enhanced by a lighter $H$ Higgs.


\begin{figure}[t!]
    \begin{center}
\centerline{
       \epsfig{file=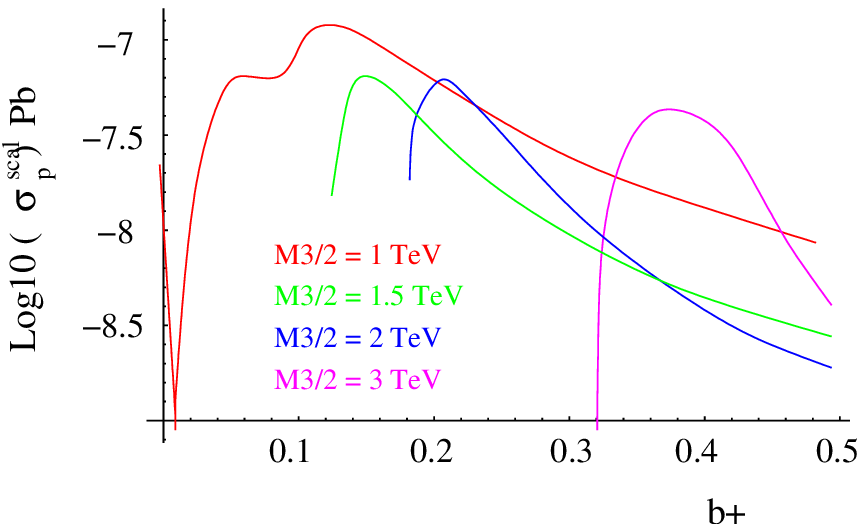,width=0.5\textwidth}
       \epsfig{file=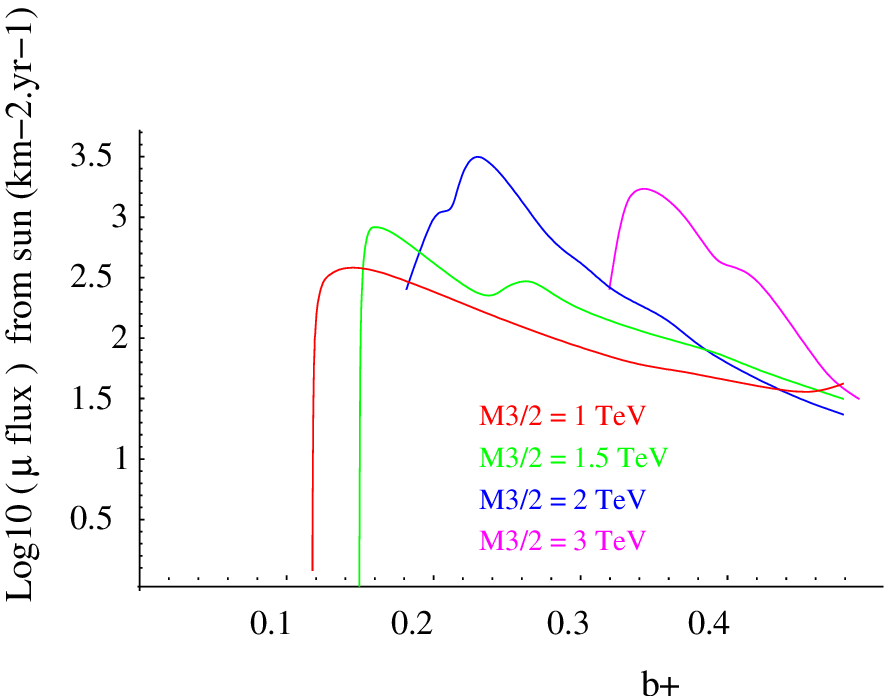,width=0.5\textwidth}}
          \caption{{\footnotesize {\bf The spin--independent scalar cross section (left) and the muon flux from the sun (right)  in the dilaton
          parameter space}, as a function of $b_+$ for
          tan $\beta=$ 35 and different values of $M_{3/2}$ .}}
          \label{fig:dilatonb+}
    \end{center}
\end{figure}

The interesting effect of $b_+$ on both direct ($\sigma^{scal}_{\chi- p}$) and
indirect (${\rm flux}^{\odot}_{\mu}$) detection is
illustrated on Fig. \ref{fig:dilatonb+}. We see that decreasing $b_+$ can 
give a
gain of one order of magnitude for direct detection and up to 2 orders
of magnitude for indirect detection which is favoured twice by the higgsino
fraction (capture plus annihilation in gauge bosons giving more energetic
neutrinos/muons). $\sigma^{scal}_{\chi- p}$ being proportional to 
$z_{\chi,1}z_{\chi,3(4)}$, finally decreases when $b_+(\mu)$ is further decreased
($\chi$ too much higgsino like). 
The bumps on Fig. \ref{fig:dilatonb+} (left panel) occur around  
the $\mu$-$M_1$ crossing (Fig. \ref{fig:m1m2m3mudilat}). This requires higher 
values of $b_+$ for higher values of $M_{3/2}$. The last 
re-increasing of $\sigma^{scal}_{\chi-p}$ (for $M_{3/2}=1\ {\rm TeV}$) comes 
from the $\mu$ evolution 
(Fig. \ref{fig:m1m2m3mudilat}) and $z_{\chi,1}z_{\chi,3(4)}$ dependence.

${\rm Fluxes}^{\odot}_{\mu}$ have the same behaviour with $M_{3/2}$ and $b_+$, 
but being proportional to $z^2_{\chi 3(4)}$, when $b_+$  decreases they still 
increase  or actually follow  the (inverse) evolution of the 
parameter $\mu$ as can be clearly seen by comparing Figs. 
\ref{fig:m1m2m3mudilat} and \ref{fig:dilatonb+} (right panel). 
Starting from high values of $b_+$, we can also 
notice the first bump corresponding to $m_{\chi}=m_t$ because neutrino coming 
from $\chi\chi \rightarrow t\bar{t}$ annihilation are less energetic than 
$\chi\chi \rightarrow W^+W^-,\ ZZ$. This very favourable effect on neutralino 
dark matter relic density and detection rates consisting in decreasing
$\mu$ and scalar masses through $M_3|_{GUT}$ running effect has been
pointed out in \cite{Mynonuniv} but is directly related here to the
$b_+$ parameter.

\begin{figure}[h!]
    \begin{center}
\centerline{
       \epsfig{file=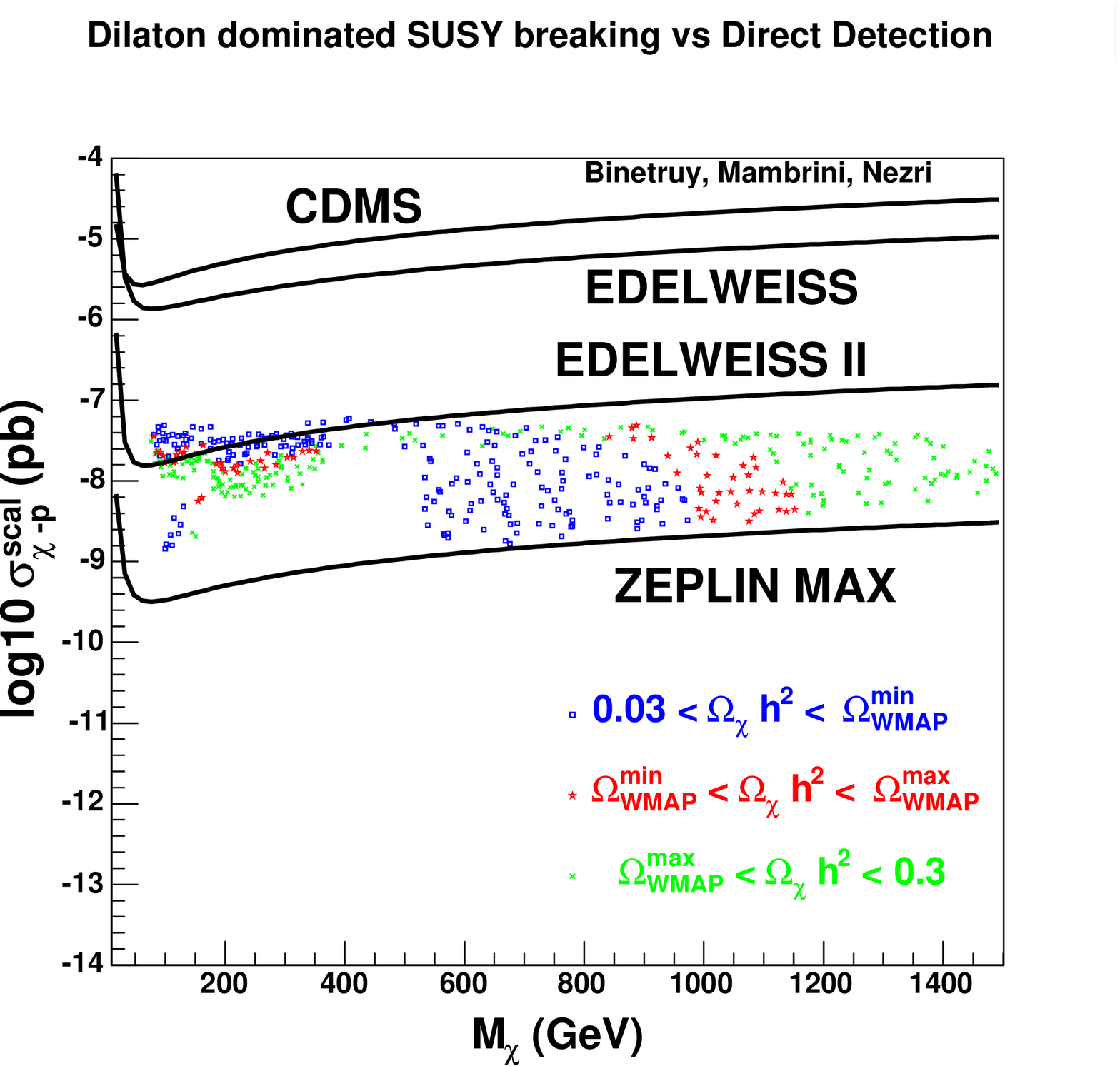,width=0.45\textwidth}
       \epsfig{file=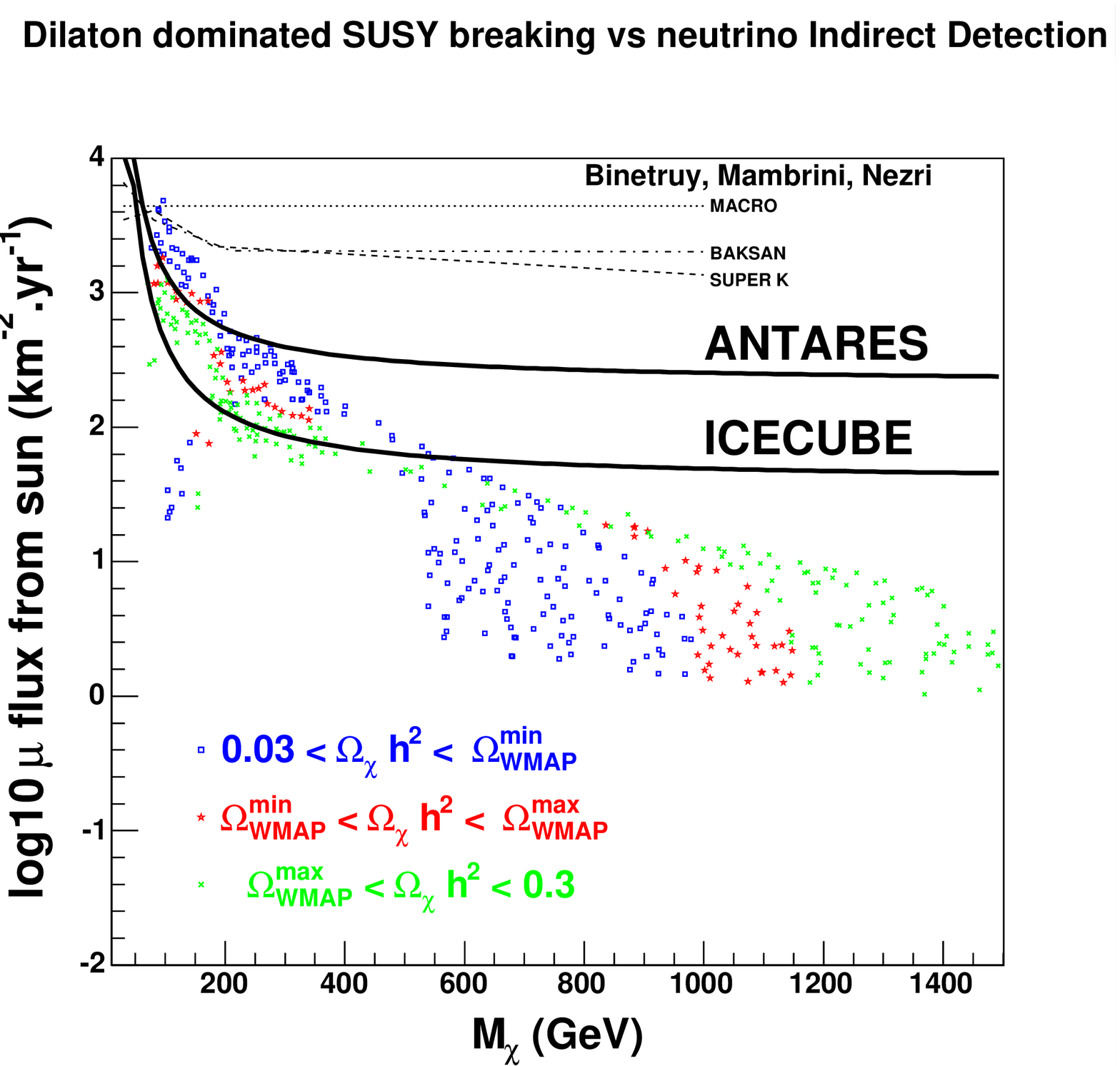,width=0.45\textwidth}}
          \caption{{\footnotesize {\bf The spin--independent scalar
          cross section (left) and the muon flux from the sun (right) 
          in the dilaton
          parameter space}, 
          as a function of the neutralino mass $M_{\chi}$ after a scan
          on $b_+$ and $M_{3/2}$ for tan $\beta=$ 5, 20 and 35.
          Accelerators and cosmological constraints (taken for $\mu > 0$)
          have been included. 
          For a description of the experimental constraints applied, see 
          Section~\ref{sec:constraints}.}}
          \label{fig:dilaton-exp}
    \end{center}
\end{figure}

To conclude this dilaton case, if we compare model predictions with
experimental sensitivities for both direct (Fig.
\ref{fig:dilaton-exp} left panel) and neutrino indirect detection
(Fig. \ref{fig:dilaton-exp} right panel), we see that dilaton
models satisfying both accelerator constraints and approximate WMAP relic
density give mixed higgsino-gaugino neutralino with generically
 high detection rates : 
$\Omega_{\chi}h^2\sim (\Omega h^2)^{WMAP}_{CDM}$,
$\sigma^{scal}_{\chi -p}\sim 10^{-(7 \ {\rm to}\ 9)}$ pb and 
${\rm flux}_{\mu}^{\odot}\sim 10^{0  \ {\rm to}\ 3} {\rm km^{-2}\ yr^{-1}}$.

One of the most interesting feature of the
dilaton type models is its closed parameter space. Indeed, the beta function
of the first condensing gauge group cannot exceed the largest one of the 
models ($b_{E8} \sim 0.57$). On the other way, the No EWSB condition forbids 
high values of $M_{3/2}$, the upper limit depending on $\tan{\beta}$. Imposing 
WMAP constraints on the relic density, in this this closed 
$(M_{3/2},b_+)$ plane, neutralino mass is limited to 500 GeV (resp. 1500 GeV) 
for $\tan{\beta}=35$ (resp. 5). Those limits can be compared to the mSUGRA 
case where $m_{\chi}<500$ GeV for $\tan{\beta}<45$\cite{EllisWmap} (small 
values of $M_0$ and $M_{1/2}$)  but where this bound can be higher ($\simeq$ TeV) if one considers the hyperbolic branch/focus point region \cite{Nath1}. 

Dilaton models  could be detected by future dark matter searches especially 
a ton-size direct detection experiment like Zeplin \cite{Zeplin}, or, for points 
with $m_{\chi} \leq 500$ GeV a ${\rm km^{3}}$ neutrino telescope like Icecube 
\cite{Ice3Edsjo}. 

\section{Conclusion}

In the specific context of the class of string models that we have considered, 
we have seen that the predictions regarding dark matter are strikingly 
different according to the type of supersymmetry breaking considered.
In the case of moduli domination, one does not expect any signal in the 
forthcoming direct or indirect (neutrino) detection experiments. On the other
hand, these experiments should not miss the neutralino signal in the case of 
dilaton domination. Thus the detection of dark matter or the absence of 
detection may give key information on the nature of supersymmetry breaking, at 
least in the context of this given class of models.

Obviously there are connections between these results and detection of the LSP 
at colliders. Small direct detection cross sections or small indirect 
detection fluxes are obviously correlated with small production cross sections 
at colliders. In any case, it is interesting for collider searches to note the 
characteristics of the regions that satisfy the criterion of satisfactory 
relic density. For moduli domination, we have identified two regions of 
interest: one where  $m_{\chi}\sim m_{\chi^+_1} \sim m_{\chi^0_2}$ 
through the  bino and wino content of the LSP (for sufficiently large values 
of $\tan \beta$), and the other one close to stau LSP region where
$m_{\tilde{\tau}_1}\sim m_{\chi}$. In the case of dilaton domination, the
cosmologically interesting region corresponds to a LSP with a proper higgsino 
content ($m_{\chi}\sim m_{\chi^+_1}\sim m_{\chi^0_2}$). Furthermore the 
parameter space being closed, this case gives an upper bound on neutralino mass: $m_{\chi}<1500$ GeV.

We think that the type of conclusions that we have reached is relevant to a larger class of models. It remains however to 
make similar analyses on other classes of models to see if
dark matter searches, whether they are positive or negative, give interesting 
indications on the way supersymmetry is broken, and on the way it is 
transmitted to the observable sector.

\vskip 1.cm

{\bf {\large Acknowledgements}}

E.N. acknowledges support from the Belgian Federal Government under contract 
IAP V/27, the French Community of Belgium (ARC) and the IISN. Y.M. would like
to thank J.L. Kneur and E. Dudas for useful discussions and comments. We would
like to thank G. Belanger for the essential help that she provided for most of
the numerical part of this work.

\vspace{1cm}

\nocite{}
\bibliography{bmn}
\bibliographystyle{unsrt}

\end{document}